\documentclass[usenatbib]{mnras}
\usepackage{graphicx} % Required for inserting images
\usepackage{acronym}
\usepackage{enumitem}
\usepackage{booktabs}
\usepackage{xcolor}
\usepackage{tikz}
\usetikzlibrary{shapes.geometric, arrows.meta, positioning, fit, backgrounds, calc}

\title[GWTC-4.0's near-threshold events]{Lurkers at the Threshold: GWTC-4.0's near-threshold events}
\author[D. Williams \& M. J. Williams]{Daniel Williams\thanks{daniel.williams@glasgow.ac.uk}$^1$ and Michael J. Williams$^{2, 3}$\\
$^1$Institute for Gravitational Research, School of Physics \& Astronomy, University of Glasgow,  G12 8QQ, United Kingdom\\
$^2$Max Planck Institute for Gravitational Physics (Albert Einstein Institute), Am M\"uhlenberg 1, 14476 Potsdam, Germany\\
$^3$Institute of Cosmology and Gravitation, University of Portsmouth, Portsmouth PO1 3FX, United Kingdom
}
\date{\today}

\newcommand{\gwtcfourcandidates}{128}
\newcommand{\gwtcfourfar}{87}
\newcommand{\gwtcfourpe}{86}
\newcommand{\inbetweencandidates}{42}
\newcommand{\pastro}{\ensuremath{p_{\mathrm{astro}}}}
\newcommand{\farthreshold}{\ensuremath{1 / \mathrm{year}}}
\newcommand{\code}[1]{\texttt{#1}}
\newcommand{\BILBY}{\code{bilby}}
\newcommand{\ASIMOV}{\code{asimov}}
\newcommand{\BAYESWAVE}{\code{bayeswave}}
\newcommand{\DYNESTY}{\code{dynesty}}
\newcommand{\PESUMMARY}{\code{pesummary}}

\newcommand{\nanalysed}{42}      % events in the asimov project
\newcommand{\napproved}{29}      % events with an approved IMRPhenomXPHM run
\newcommand{\nnoapproved}{13}    % events without one
\newcommand{\nnoise}{2}
\newcommand{\nglitch}{5}
\newcommand{\nunconstrained}{5}
\newcommand{\nmultimodal}{1}

\newcommand{\tpass}{\textcolor{black!35}{$\circ$}}
\newcommand{\twarn}{\textcolor{orange!85!black}{\textbf{W}}}
\newcommand{\tfail}{\textcolor{red!75!black}{\textbf{F}}}

\newacro{GWOSC}[GWOSC]{gravitational-wave open science centre}
\newacro{far}[FAR]{false-alarm rate}
\newacro{gw}[GW]{gravitational wave}
\newacro{GWTC5}[GWTC-5.0]{the fifth revision of the gravitational-wave transient catalog}
\newacro{GWTC4}[GWTC-4.0]{the fourth revision of the gravitational-wave transient catalog}
\newacro{GWTC3}[GWTC-3]{the third revision of the gravitational-wave transient catalog}
\newacro{O4a}[O4a]{the first part of the fourth observing run}
\newacro{O4b}[O4b]{the second part of the fourth observing run}
\newacro{O4c}[O4c]{the third part of the fourth observing run}
\newacro{O3}[O3]{the third observing run}
\newacro{PE}[PE]{parameter estimation}
\newacro{PSD}[PSD]{power spectral density}
\newacro{SNR}[SNR]{signal-to-noise ratio}
\newacro{DQ}[DQ]{data quality}
\newacro{BBH}[BBH]{binary black hole}
\newacro{NSBH}[NSBH]{neutron star-black hole binary}
\newacro{NS}[NS]{neutron star}

\tikzset{
  >={Stealth[length=2mm]},
  wfstep/.style={rectangle, rounded corners=2pt, draw=black!70, fill=white, align=center,
               font=\footnotesize, minimum height=8mm, text width=2.6cm, inner sep=3pt},
  wfauto/.style={wfstep, fill=blue!8},          % performed automatically by asimov
  wfmanual/.style={wfstep, fill=orange!15},     % performed by an analyst
  wfdecision/.style={diamond, aspect=2.2, draw=black!70, fill=yellow!15, align=center,
                   font=\footnotesize, inner sep=1pt, text width=2.2cm},
  wfoutcome/.style={rectangle, draw=black!70, fill=green!10, align=center,
                  font=\footnotesize, minimum height=8mm, text width=2.4cm, inner sep=3pt},
  wfreject/.style={wfoutcome, fill=red!10},
  flow/.style={->, thick, black!70},
  lbl/.style={font=\scriptsize, inner sep=1pt},
}

\begin{document}

\maketitle

\begin{abstract}

We present the first systematic parameter-estimation analysis of 42 candidate gravitational-wave signals identified in GWTC-4.0 with a probability of astrophysical origin $p_\mathrm{astro} \geq 0.5$, but a false-alarm rate of at least $1\,\mathrm{yr}^{-1}$, which were therefore not selected for parameter estimation in the catalogue. We use a catalogue-style analysis workflow and publicly available strain data to obtain posterior samples for these candidates, of which 29 pass our validation criteria. Of the remaining 13 candidates, seven are consistent with noise or impacted by instrumental artefacts, while six yield results that cannot be reliably interpreted.

We identify several interesting candidates: a possible neutron star--black hole merger, GW230920\_064709, and two candidates with evidence for non-zero effective spin: GW230902\_172430 and GW230823\_142524. The latter also shows mild evidence of precession, although further analysis is needed to establish the robustness of this result. We quantify the information gained from these observations using the Kullback--Leibler divergence, finding that the inferred spin parameters are generally only weakly constrained.

We present a reproducible analysis workflow that uses publicly available data and software, without access to LVK collaboration resources, and incorporates automated posterior-validation tools. We also provide a data release containing results in a format compatible with those from GWTC-4.0.

Our results demonstrate that useful parameter constraints can be obtained for many candidates below the catalogue's parameter-estimation significance threshold, while highlighting the challenges of analysing marginal signals and the importance of careful validation.

% \todo{Abstract: scope (candidates with $\pastro \geq 0.5$ but \ac{far} above the \ac{GWTC4} parameter-estimation threshold), method (a GWTC-like \ac{PE} workflow with a single waveform, run on public data), headline outcome (\napproved{} of \nanalysed{} events yielded an approved analysis), caveats (prior-dominated and glitch-affected events), and the data release.}
\end{abstract}


\section{Introduction}  

The first results from the fourth observing run of the LIGO, Virgo, and KAGRA detectors~\citep{ligoscientificcollaborationAdvancedLIGO2015, acerneseAdvancedVirgoSecondgeneration2015,2021PTEP.2021eA101A} were revealed in \ac{GWTC4}~\citep{catalog-gwtc-4d0}
The associated release of strain data from \ac{O4a} of the LIGO, Virgo, and KAGRA detectors has provided a rich new collection of signals from coalescing compact binaries.
\gwtcfourcandidates{} new candidates were identified in the data, and results from detailed parameter estimation were provided for \gwtcfourpe{} of these candidates.
This discrepancy arises from the choice to define two different thresholds when compiling the catalogue.
The first is the probability of signals being astrophysical, \pastro{}.
This was introduced in \citet{abbott_rate_2016}, and is used to define the larger of the two sets.
In the previous iteration of the transient catalogue this was the only criterion which was used.
However, in an attempt to reduce the number of signals which might be in some way non-astrophysical in origin, \ac{GWTC4} uses a second threshold for the candidates where the labour-intensive and computationally expensive process of parameter estimation is conducted: the signals' \ac{far}.

The \pastro{} threshold determines whether a signal is included in the list of new candidates, with an identified signal classed as a candidate if $\pastro \geq 0.5$ in at least one of the search pipelines.
\gwtcfourcandidates{} candidates are identified which exceed this threshold in \ac{O4a}.
The FAR threshold applied in \ac{GWTC4} is a rate of \farthreshold{}~\citep{gwtc4_methods_2026}. % The search summary table lists 129 candidates with p_astro >= 0.5 in at least one pipeline: 87 with FAR below 1/yr and 42 at or above it, which are the events analysed here. The combined listing, which gives the p_astro of the pipeline with the lowest FAR, has 128 and 41, and omits GW230625_211655 (0.11 in gstlal, 0.61 in pycbc). None has FAR below 1/yr and p_astro < 0.5.
\gwtcfourfar{} candidates exceed the \ac{far} threshold, and parameter estimation results are presented for \gwtcfourpe{} of them in \ac{GWTC4}, the remainder being GW230630\_070659, which was determined to be of instrumental origin~\citep{catalog-gwtc-4d0}.
There were no candidates which exceeded the \ac{far} but not the \pastro{} threshold.

In this work we examine the \inbetweencandidates{} candidates which lie between these two thresholds, for which $\pastro \geq 0.5$ but \ac{far} $\geq \farthreshold{}$.
These candidates are, by construction, the least significant candidates which the catalogue considers more likely astrophysical than not, and so they are the signals for which the question ``what do we learn from the data?'' is most sensitive to the details of the analysis.
Their signals are weak, their posterior distributions are often dominated by the prior, and, since they were not examined as thoroughly as the candidates in the main catalogue, some are likely to be affected by non-Gaussian noise.

We closely follow the approach of \citet{catalog-beyond}, which used the \ASIMOV{} workflow manager~\citep{williams_asimov_2023} to reproduce a catalogue-style \ac{PE} analysis for candidates reported in the literature but absent from \ac{GWTC3}.
Here we apply the same philosophy to the candidates which lie in the gap between the two thresholds of \ac{GWTC4}, and we use analysis configurations which are comparable to those used for the parameter estimation results presented in~\citet{catalog-gwtc-4d0}.

In section~\ref{sec:eventselection} we discuss the method used to select the candidate events (candidates) which we analyse; in section~\ref{sec:pe} we discuss the method used for estimating the source parameters for these candidates, and to verify the efficacy of these analyses. In section~\ref{sec:outcomes} we discuss the initial outcomes of these analyses, specifically whether the candidates provide results which are consistent with being of astrophysical origin. 
In section~\ref{sec:sourceparameters} we discuss the inferred source parameters for those signals which are consistent with being of astrophysical origin, and we provide discussion of these candidates.

\section{Candidate Selection}\label{sec:eventselection}

The candidates which were selected for this work are drawn from Table 2 of~\citet{catalog-gwtc-4d0}, which lists all of the new candidates identified in the \ac{O4a} observing run which were determined to have a \pastro{} value greater than or equal to 0.5, but a \ac{far} value greater than or equal to \farthreshold{}.
We take the event times, the detectors in which each candidate was identified, and a search estimate of the chirp mass from the \ac{GWTC4} search data release~\citep{gwtc4-search-datarelease}.

\begin{figure}
    \centering
    \includegraphics[width=\columnwidth]{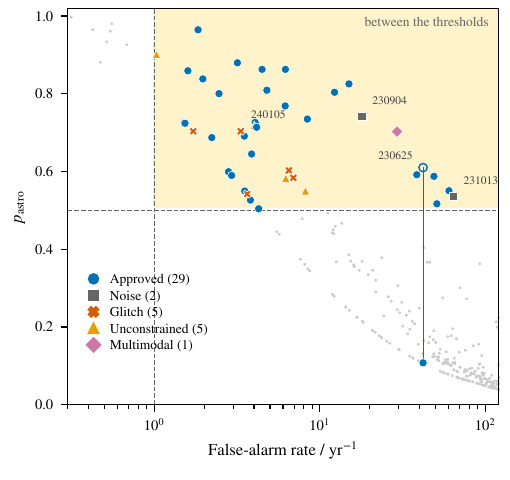}
    \caption{The false-alarm rate and $\pastro$ of every trigger in the \ac{GWTC4} search summary (grey). The $\pastro$ shown is that of the combined listing of the search summary, which is that of the pipeline with the lowest \ac{far}, and triggers with $\pastro < 0.5$ in every pipeline are not candidates. GW230625\_211655 is a candidate because its $\pastro$ is 0.61 in \texttt{pycbc}, shown by the open marker, although it is 0.11 in the combined listing (Figure~\ref{fig:pipelines}). Of the \gwtcfourcandidates{} candidates, the \gwtcfourfar{} with a false-alarm rate below \farthreshold{} (top left, of which 80 lie beyond the left edge of the plot) were analysed in \ac{GWTC4}. The plot is cut off at a false-alarm rate of 120 per year, and a further 1034 triggers, all with $\pastro < 0.5$, lie beyond the right edge. The \inbetweencandidates{} candidates in the shaded region, with $\pastro \geq 0.5$ and a false-alarm rate of at least \farthreshold{}, are the candidates analysed in this work; they are coloured by their outcome (Section~\ref{sec:outcomes}), and the number of candidates with each outcome is given in the legend.}
    \label{fig:selection}
\end{figure}

We show an overview of the candidates in Figure~\ref{fig:selection}, in the context of both the least significant candidates for which parameter estimation was provided in \ac{GWTC4} (upper left corner) and the most significant candidates for which we do not provide parameter estimation (lower right corner).
The probability that a candidate is of terrestrial origin is $1 - \pastro$, so summing the \inbetweencandidates{} candidates' probability of terrestrial origin gives the number of them which we expect not to be astrophysical, which is 13.4, using the \pastro{} of the combined listing of the search summary, which is that of the pipeline with the lowest \ac{far}.
A substantial fraction of the candidates which we analyse are therefore expected to be noise or instrumental artefacts, and we return to this in Section~\ref{sec:outcomes}.

\begin{figure*}
    \centering
    \includegraphics[width=0.8\linewidth]{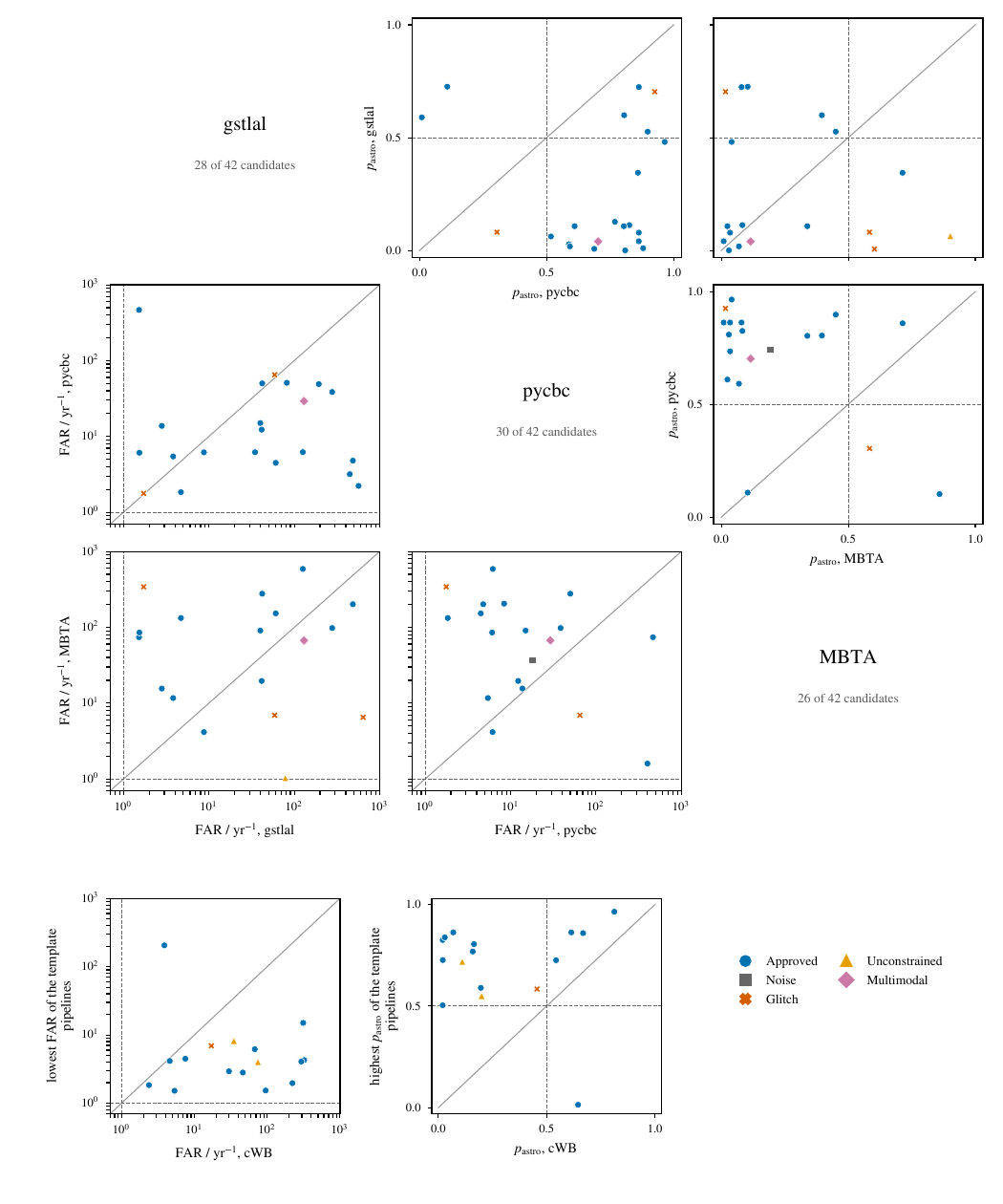}
    \caption{The \ac{far} (lower triangle) and $\pastro$ (upper triangle) which each pair of the template-based search pipelines gives to the candidates found by both, and (bottom row) the values of cWB against the lowest \ac{far} and the highest $\pastro$ of the template-based pipelines, for the 16 candidates which were found by both cWB and a template-based pipeline. Dashed lines show the thresholds of one per year and $\pastro = 0.5$, and the solid line shows equal values. The candidates are coloured by their outcome (Section~\ref{sec:outcomes}).}
    \label{fig:pipelines}
\end{figure*}

The search pipelines can differ substantially in their assessment of the same candidate, as we show in figure~\ref{fig:pipelines}.
Of the \nanalysed{} candidates, 33 were found by more than one pipeline, and none of these has $\pastro \geq 0.5$ in every pipeline which found it.
The \acp{far} of the 27 candidates which were found by more than one of the template-based pipelines (\texttt{gstlal}~\citealp{software-gstlal}, \texttt{pycbc}~\citealp{software-pycbc}, and \texttt{MBTA}~\citealp{software-mbta}) differ between the pipelines by a median factor of 24, and by as much as a factor of 600.
Taken alone, \texttt{pycbc} gives $\pastro \geq 0.5$ to 25 of the 30 candidates which it found, but \texttt{gstlal} and \texttt{MBTA} do so for only 10 of 28 and 10 of 26, and cWB~\citep{software-cwb} for 6 of 17, so that the list of candidates depends on the pipeline used, and we use the union of them.

In the combined listing in~\citep{gwtc4-search-datarelease}, the $\pastro$ stated for each candidate, the $\pastro$ from the pipeline which gave it the lowest \ac{far}.
We instead select a candidate if $\pastro \geq 0.5$ in any pipeline which found it.
The two definitions agree for all candidates but one, GW230625\_211655.
\texttt{gstlal} gives it the lowest \ac{far} and a $\pastro$ of 0.11, which is therefore its value in the combined listing, but \texttt{pycbc} gives it a $\pastro$ of 0.61.
It is the only candidate which we include as a result of this choice.

\section{Parameter Estimation}\label{sec:pe}

We largely repeat the approach described in \citet{catalog-beyond}, which attempted to replicate the analysis configuration used for \ac{PE} analyses in \ac{GWTC3}~\citep{catalog-gwtc3} for a number of candidates which had been identified elsewhere in the literature.
As in our previous publication we provide a reduced set of analyses relative to what is made available in, for example, \ac{GWTC4}, and we only produce posterior samples using a single waveform model, IMRPhenomXPHM-SpinTaylor.
Each analysis is managed by \ASIMOV{}~\citep{williams_asimov_2023}, which tracks the state of every event and analysis in a ledger, and which is configured by ``blueprint'' files which are released alongside this work~\citep{datarelease}.

\subsection{Workflow}\label{sec:workflow}

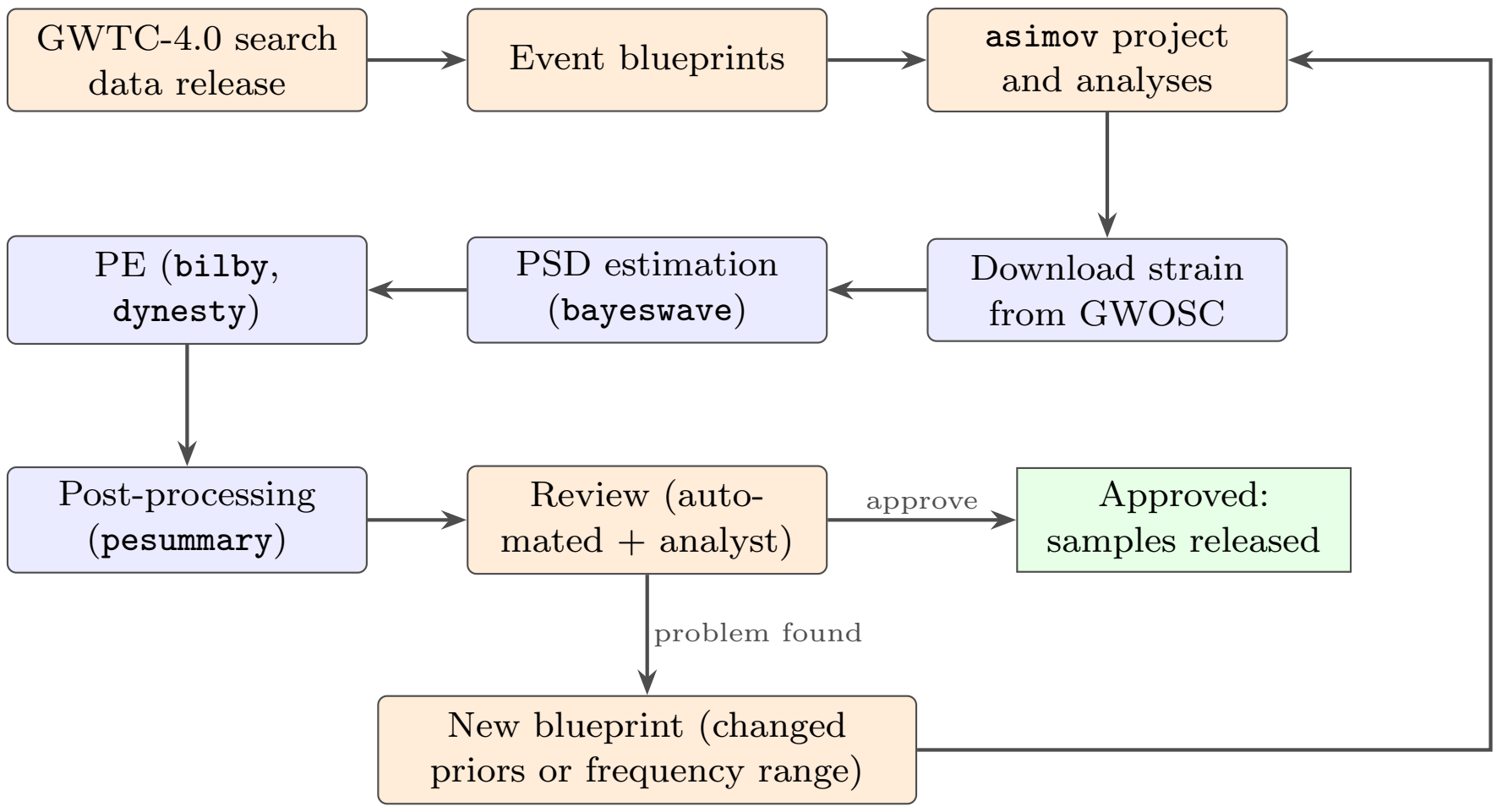
\begin{figure*}
    \centering
    \begin{tikzpicture}
        \node[wfmanual] (gwtc) at (0,0) {\ac{GWTC4} search data release};
        \node[wfmanual] (bp) at (3.6,0) {Event blueprints};
        \node[wfmanual] (proj) at (7.2,0) {\ASIMOV{} project and analyses};
        \node[wfauto] (data) at (7.2,-1.8) {Download strain from \ac{GWOSC}};
        \node[wfauto] (psd) at (3.6,-1.8) {\ac{PSD} estimation (\BAYESWAVE{})};
        \node[wfauto] (pe) at (0,-1.8) {\ac{PE} (\BILBY{}, \DYNESTY{})};
        \node[wfauto] (post) at (0,-3.6) {Post-processing (\PESUMMARY{})};
        \node[wfmanual] (rev) at (3.6,-3.6) {Review (automated + analyst)};
        \node[wfoutcome] (out) at (7.8,-3.6) {Approved: samples released};
        \node[wfmanual, text width=4cm] (new) at (3.6,-5.4) {New blueprint (changed priors or frequency range)};

        \draw[flow] (gwtc) -- (bp);
        \draw[flow] (bp) -- (proj);
        \draw[flow] (proj) -- (data);
        \draw[flow] (data) -- (psd);
        \draw[flow] (psd) -- (pe);
        \draw[flow] (pe) -- (post);
        \draw[flow] (post) -- (rev);
        \draw[flow] (rev) -- node[lbl, above] {approve} (out);
        \draw[flow] (rev) -- node[lbl, right] {problem found} (new);
        \draw[flow] (new.east) -- (10.2,-5.4) -- (10.2,0) -- (proj.east);
    \end{tikzpicture}
    \caption{An overview of the analysis workflow used for this work. 
    Orange boxes are steps which required the input of a human analyst; blue boxes are performed automatically by \ASIMOV{}. 
    When a problem was identified when reviewing results, a new analysis blueprint was prepared to correct the problem and the automatic steps were then repeated.}
    \label{fig:workflow}
\end{figure*}

The workflow we employed for this work is shown in Figure~\ref{fig:workflow}, and, has six steps.
The first two steps were performed outwith the asimov automation framework, in order to prepare the analysis:
\begin{enumerate}
    \item we extracted event information from the search candidate list contained in the \ac{GWTC4} search data release~\citep{gwtc4-search-datarelease}. We then prepared a blueprint file for each candidate, and used these to add the information about the candidate to asimov;
    \item we then added the required analyses to each event in the project.
\end{enumerate}
The remaining steps are then performed automatically, for each candidate, by \ASIMOV{}:
\begin{enumerate}[resume]
    \item strain data, stored in GWF frame files~\citep{LIGOT970130v3SpecificationCommon}, were downloaded from the \ac{GWOSC};
    \item noise estimation was performed using the \BAYESWAVE{}~\citep{cornish_bayeswave_2015} software, which fits and removes the signal from the data to produce an estimate of the \ac{PSD} from the noise alone (an ``on-source'' estimate; as was also done for \ac{GWTC4});
    \item \ac{PE} was performed using the \BILBY{}~\citep{code-bilby} library with the \DYNESTY{}~\citep{speagle_dynesty_2020} nested sampling algorithm; and
    \item post-processing was performed on the posterior samples to create derived parameters, data in an interchangeable format, and summary pages with \PESUMMARY{}~\citep{hoy_pesummary_2021}.
\end{enumerate}
We then utilised a mixture of automated tools and expert inspection to review the results of each analysis (see Section~\ref{sec:review}) in order to determine if the analysis could be accepted, if it needed to be rerun with altered settings, or if the \ac{PE} implied that the signal was unlikely to be of astrophysical origin.

\subsection{Analysis configuration}\label{sec:config}

All of the analyses use the IMRPhenomXPHM waveform~\citep{pratten_xphm_2021} in its ``SpinTaylor'' precession configuration~\citep{colleoni_spintaylor_2024}.
We choose a reference frequency of 20\,Hz, and with the \texttt{PhenomXHMReleaseVersion}, \texttt{PhenomXPFinalSpinMod}, and \texttt{PhenomXPrecVersion} settings fixed to 122022, 2, and 320 respectively.
We chose this waveform as, in general, out of the four waveforms which were presented in \ac{GWTC4} it has a combination of broad applicability across the parameter space, and the most modest computational requirements for evaluation of the waveforms used in \ac{GWTC4}.

We set a segment duration (4, 8, or 64\,s) and a sampling frequency (between 256 and 8192\,Hz) individually for each candidate, using the point estimate of the chirp mass from the search to inform this choice.
To account for windowing roll-off in the likelihood function, the maximum frequency used in the likelihood is set to 87.5\% of the Nyquist frequency for the data, and the analysed data segment is then generated so that it ends 2\,s after the arrival time of the signal as estimated in the search results.
The minimum frequency used in the likelihood was 20\,Hz for every detector in every successful analysis, however we tested alternative minimum frequencies for some candidates which showed evidence of glitches in the data affecting inference in an attempt to exclude the glitch from the analysis (this will be discussed in greater detail in section~\ref{sec:outcomes}).

We estimated the noise \ac{PSD} of the data from each detector using \BAYESWAVE{}~\citep{cornish_bayeswave_2015, cornish_bayeswave_2021} which is capable of performing ``on-source'' \ac{PSD} estimation by first fitting-out signal power using wavelets (Section~\ref{sec:workflow}).
This is performed separately for each signal, as noise in the detector varies over the period of an observing run. 

We then performed Bayesian inference using the \BILBY{}~\citep{code-bilby, romeroshaw_bilbypipe_2020} framework.
We used the \DYNESTY{}~\citep{speagle_dynesty_2020} algorithm to perform nested sampling of the posterior probability distribution, with 1000 live points and the acceptance-walk proposal with 60 accepted steps.
We combined samples from three parallel sampling jobs in order to increase the number of final samples produced.
We marginalised the likelihood over the luminosity distance parameter, enabling the sampling to be performed more efficiently; the luminosity distance can then be reconstructed in post-processing to provide samples for this parameter \emph{post hoc}. 
These are the same settings and processes used in the production of samples for parameter estimation produced by \BILBY{} in both \ac{GWTC4}~\citep{catalog-gwtc-4d0} and \ac{GWTC5}~\citep{catalog-gwtc-5}.
In common with \ac{GWTC4}, we did not integrate over the phase or the arrival time.
\ac{GWTC4} also marginalises over the uncertainty in the detector calibration; similarly to \citet{catalog-beyond} we do not consider the effects of detector calibration uncertainty, which we do not expect to introduce any substantial difference to the posterior distributions for low \ac{SNR} signals.

We used the same prior morphologies as those used in \ac{GWTC4}; priors on component masses are uniform in the detector-frame, though these are implemented as priors on the chirp mass and the mass ratio, $q$, with $0.05 \leq q \leq 1$ and component masses between 1 and 1000\,\ensuremath{M_\odot}.
The lower bound on the mass-ratio is determined by the range over which the calibration of the IMRPhenomXPHM waveform approximant is valid.
The priors on spin magnitudes are uniform between 0 and 0.99, with isotropic orientations.
We set a prior on luminosity distance which is uniform in comoving volume and source-frame time between 10 and 20\,000\,Mpc.

The bounds on the chirp mass are set for each candidate (in some cases after successive analyses, see section~\ref{sec:review}).
We did not extend the upper bound above 160\,\ensuremath{M_\odot}, because the waveform will have minimal power in the detector's sensitive frequency range. 

We calculated source-frame quantities assuming the Planck15 cosmology~\citep{planck_2015_xiii}, as implemented in \texttt{LALSuite}~\footnote{N.B. that in common with other LVK analyses we use $H_0=67.90\,{\rm km/Mpc/s},\ \Omega_{\rm m}=0.3065$}.

\subsection{Differences from the \ac{GWTC4} analysis}\label{sec:differences}

Our aim in preparing this work has been, wherever possible, to use identical processes and configurations to those used when constructing transient catalogues such as \ac{GWTC4}~\citep{gwtc4_methods_2026}.
However, as we did not perform this analysis on LVK Collaboration infrastructure, there are a number of differences, which are mostly subtle or inconsequential:

\begin{itemize}
    \item \textbf{Data retrieval.} We used data from \ac{GWOSC}~\citep{gwosc_o4a_2025}. 
    These were accessed using different infrastructure compared to the data used by the collaborations during their proprietary period.
    The data prepared for public release are identical to those used internally within the collaboration, but metadata labels used in internal and published data products are different, and so a blueprint intended for use on LVK infrastructure must be updated for use on public data.
    Using \ac{GWOSC} data requires an additional preliminary step (provided by the \texttt{asimov-gwdata} python package~\footnote{https://pypi.org/project/asimov-gwdata/} in the workflow to retrieve the strain data.
    This is not required when running on collaboration infrastructure, as both \BAYESWAVE{} and \BILBY{} are capable of retrieving proprietary data automatically.
    While this alters the workflow, it has no effect on the analysis itself.
    
    \item \textbf{Prior bounds.} In conventional catalogue analyses the prior bounds on chirp mass and luminosity distance are constrained as tightly as possible in order to reduce the prior volume which must be explored during sampling.
    To do this, posterior samples created during a preliminary ``online'' \ac{PE} step are typically used to inform the bounds used for the ``production'' analysis.
    These preliminary analyses are not released to the public, and so we set broad bounds at the outset, informed by the search estimate of the chirp mass (which for some candidates is a poor guide: see the discussion of Figure~\ref{fig:searchpe}), and rerun the analysis with broader priors in situations where railing is identified (see Section~\ref{sec:review}).
    
    \item \textbf{Waveform.} We performed analysis with only one waveform, so we do not provide combined samples from multiple waveforms. 
    As a consequence we are unable to provide any analysis of the effect of waveform systematics in our analyses. 
    
    \item \textbf{Software.} All but one of the approved analyses used \BILBY{} 2.6.0, with \texttt{bilby\_pipe} 1.4.0, \DYNESTY{} 2.1.5, and \texttt{LALSuite} 7.25, and the on-source noise power spectral densities were estimated with \BAYESWAVE{} 1.1.1. 
    The analysis of GW230902\_122814, and the final analyses of GW230925\_143957 and GW231002\_143916, which we rejected, were run later in a new software environment, with \BILBY{} 2.8.2, a development version of \texttt{bilby\_pipe} (1.10.0.dev13), \DYNESTY{} 3.1.0, \texttt{LALSuite} 7.25, and \BAYESWAVE{} 1.1.1.
    
    \item \textbf{Calibration}. The uncertainty in detector calibration is not considered in our analyses.
\end{itemize}

\subsection{Review and rerun procedure}\label{sec:review}

In the absence of preliminary analyses to inform our choice of prior bounds, most of our candidates required more than one analysis before we obtained a valid set of posterior probability samples.
In many cases it is possible to identify and diagnose problems in a posterior distribution by examining histograms of the marginal posterior samples.
However, the scale of inference required to complete a catalogue increases with each new observing run.
Even with a large group of analysts review becomes rapidly impractical and error-prone, and this is exacerbated for small-scale projects such as this one.
To mitigate this, we have developed a set of automated checks which can be used to flag common problems, and these are described in Appendix~\ref{sec:posterior-tools}.
Analyses which are flagged by these checks can then be scrutinised in greater detail by an analyst.

We checked each completed analysis for the following conditions; the first four automatically by the tests listed in Appendix~\ref{sec:posterior-tools}:
\begin{itemize}
    \item \textbf{Prior railing.} We check whether the posterior of the chirp mass or the luminosity distance is cut off by a prior bound, which we take to be the case if the bin at that end of a histogram of the samples holds more than 2\% of the density of the fullest bin.
    The upper end of the mass ratio is allowed to rail, since equal masses ($q = 1$) is a genuine boundary.
    In analyses for this project the most common case of railing was against the upper bound on chirp mass.

    \item \textbf{Multimodality.} We check whether the marginal posteriors of the mass ratio, chirp mass, and distance have more than one significant mode, which may indicate the presence of a signal which is not well explained by the signal hypothesis (for example, a glitch).
    We expect bimodality in the inclination distribution between $\theta_{JN}$ and $\pi - \theta_{JN}$ for a two-detector network, and we do not flag this as a problem. 
    Bimodality is not automatically a reason for rejecting \ac{PE} results, but it does require further investigation by an analyst if it is present in the posterior samples.
    
    \item \textbf{Signal plausibility.} We check if the data support a signal at all using the signal/noise Bayes factor and the network \ac{SNR}.
    A signal which is recovered as being far louder than a candidate below the significance threshold should be treated as suspicious, as is a candidate for which the \ac{SNR} is much higher in one detector than the others.
    
    \item \textbf{Spins.} We check if the spin posteriors rail against their bounds relative to the prior. Extremal spins shorten the waveform, producing a model which may better fit some classes of glitch.

    \item \textbf{Data quality.} We visually inspected the Q-transform of the analysed data segment for each candidate to identify common data quality problems such as scattering noise.
\end{itemize}

The last check, for data quality, is substantially less thorough than the procedure performed for large-scale catalogues such as \ac{GWTC4}; this process is labour intensive, and beyond the resources available to the authors for this work.
As a result we have chosen to be conservative; if we had suspicions of data quality issues which we could not easily resolve (for example by reducing the size of the analysis segment, or reducing the range of frequencies used in the likelihood to exclude a glitch) we rejected the candidate.
We did not perform deglitching of any data segments.

When we identified railing in chirp mass we created a new analysis blueprint with expanded prior bounds and repeated the analysis~\footnote{This was such a routine occurrence that the tool which identifies railing on chirp mass or luminosity distance is capable of automating the work required to set up this repeat analysis}.
If widening the prior does not produce a unimodal, bounded, posterior, but one which continues to become wider, the chirp mass is not being constrained by the data.
This typically occurs with high-mass, low-\ac{SNR} signal whose merger lies close to the minimum frequency of the analysis.
Further widening the prior cannot improve the quality of analysis on such a candidate, and we do not provide posterior samples for the candidates which showed this behaviour.

\section{Outcomes}\label{sec:outcomes}

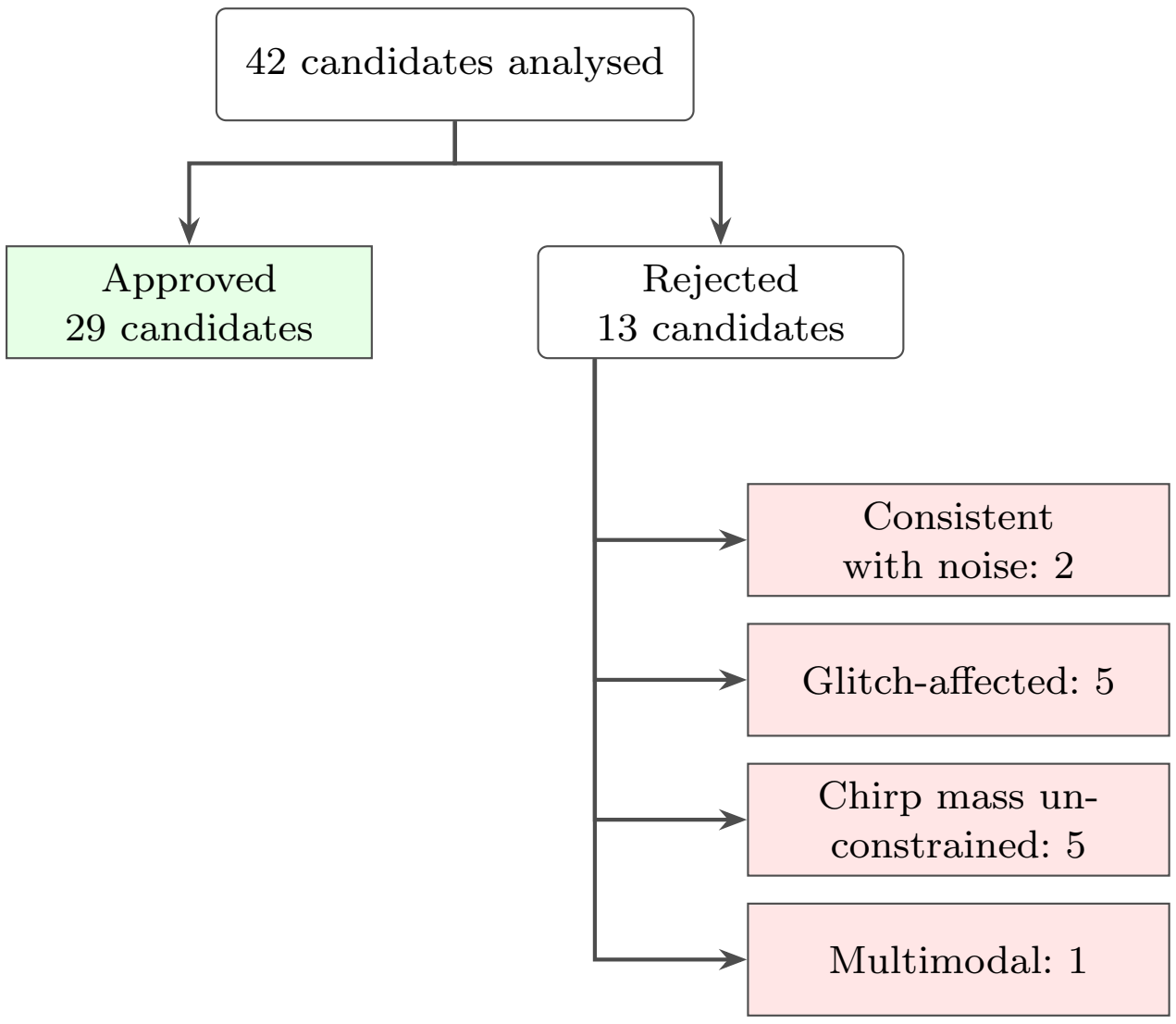
\begin{figure}
    \centering
    \begin{tikzpicture}
        \node[wfstep, text width=3.2cm] (all) at (0,0) {\nanalysed{} candidates analysed};
        \node[wfoutcome, text width=2.4cm] (ok) at (-1.9,-1.7) {Approved\\ \napproved{} candidates};
        \node[wfstep, text width=2.4cm] (no) at (1.9,-1.7) {Rejected\\ \nnoapproved{} candidates};
        \node[wfreject, text width=2.8cm] (c1) at (3.6,-3.4) {Consistent with noise: \nnoise{}};
        \node[wfreject, text width=2.8cm] (c2) at (3.6,-4.4) {Glitch-affected: \nglitch{}};
        \node[wfreject, text width=2.8cm] (c3) at (3.6,-5.4) {Chirp mass unconstrained: \nunconstrained{}};
        \node[wfreject, text width=2.8cm] (c4) at (3.6,-6.4) {Multimodal: \nmultimodal{}};

        \draw[flow] (all.south) -- ++(0,-3mm) -| (ok.north);
        \draw[flow] (all.south) -- ++(0,-3mm) -| (no.north);
        \foreach \c in {c1,c2,c3,c4} {
            \draw[flow] ([xshift=-0.9cm]no.south) |- (\c.west);
        }
    \end{tikzpicture}
    \caption{The fate of each candidate analysed.}
    \label{fig:outcomes}
\end{figure}

Of the \nanalysed{} candidates we analysed, we were able to produce posterior probability samples (which we term ``approved'') for \napproved{} and we were unable to produce acceptable samples for \nnoapproved{} (which we term ``rejected'').
The determination for each candidate is summarised in Figure~\ref{fig:outcomes}, and the candidates without an approved analysis are listed in Table~\ref{tab:problem} with further explanation for their rejection.

At least one automated check flags each of the rejected candidates (see Table~\ref{tab:matrix}), but these also flag some candidates which were ultimately approved: 7 out of the \napproved{} approved candidates are flagged by at least one check, so a flag marks an analysis for inspection and does not by itself determine that it will be rejected.
The candidates which were flagged but ultimately accepted typically had very minor railing against prior bounds or very minor secondary modes in a marginal posterior; the tests are designed to be conservative, and we made the decision to design these to produce false positives rather than false negatives.
% The counts in this paragraph are those printed by scripts/test_matrix.py

The search summary lists the pipelines which found each candidate, and Table~\ref{tab:pipelines} shows the outcomes by these.
The candidates which were found by only one pipeline were rejected more often than the others: five of the nine (56\%), compared to 8 of the other 33 (24\%), and the fraction falls from 33\% for those found by two pipelines to 12\% for those found by four.
This is a trend and not a firm result, as the sample is small and we chose the comparison after seeing the outcomes: the difference between one pipeline and several is not significant by itself (a Fisher exact test gives $p = 0.11$), and the correlation of rejection with the number of pipelines has a Kendall $\tau$ of $-0.29$ ($p = 0.042$).
Of the seven candidates which we found to be noise or glitch-affected, three (GW231013\_135504, GW231204\_090648, and GW240105\_151143) were found by a single template-based pipeline.
Which pipeline reported the lowest \ac{far} shows less of a pattern, except that five of the ten candidates for which this was MBTA were rejected; four of these are among the candidates for which MBTA used a template at the edge of its bank (see below).
Neither \ac{far} nor $\pastro$ separates the outcomes cleanly: the median \ac{far} is 4.1\,yr$^{-1}$ for the approved candidates and 6.2\,yr$^{-1}$ for the rejected ones, and we rejected 12 of the 31 candidates with a combined $\pastro < 0.8$, compared to one of the 11 (GW231002\_143916) with $\pastro \geq 0.8$.

\begin{table}
    \centering
    \begin{tabular}{lcccc}
\toprule
& Candidates & Approved & Rejected & Fraction rejected \\
\midrule
\multicolumn{5}{l}{\emph{Pipelines which found the candidate}} \\
\quad 1 & 9 & 4 & 5 & 56\% \\
\quad 2 & 15 & 10 & 5 & 33\% \\
\quad 3 & 10 & 8 & 2 & 20\% \\
\quad 4 & 8 & 7 & 1 & 12\% \\
\midrule
\multicolumn{5}{l}{\emph{Pipeline with the lowest FAR}} \\
\quad pycbc & 19 & 15 & 4 & 21\% \\
\quad gstlal & 11 & 8 & 3 & 27\% \\
\quad MBTA & 10 & 5 & 5 & 50\% \\
\quad cWB & 2 & 1 & 1 & 50\% \\
\midrule
\multicolumn{5}{l}{\emph{$\pastro$}} \\
\quad $< 0.8$ & 31 & 19 & 12 & 39\% \\
\quad $\geq 0.8$ & 11 & 10 & 1 & 9\% \\
\bottomrule
\end{tabular}

    \caption{The outcomes of the candidates by the pipelines which found them, by the pipeline which reported the lowest \ac{far}, and by $\pastro$.}
    \label{tab:pipelines}
\end{table}

The \pastro{} values also tell us how to read the outcomes.
The expected number of terrestrial candidates is 13.4, similar to the 13 which we rejected, but they are not the same candidates: 9.0 of the expected terrestrial candidates are among the 29 which we approved, and only 4.4 among the 13 which we rejected (2.6 among the seven consistent with noise or glitch-affected).
Rejection means that we could not obtain a reliable and informative \ac{PE} result, and not that the candidate is terrestrial; equally, if the values of $\pastro$ are well calibrated, a number of the approved candidates will be terrestrial in origin, but we cannot identify them as such.

\begin{table*}
    \centering
    \caption{The \nnoapproved{} candidates for which we did not produce an ``approved'' set of posterior samples, with the simple ``Outcome'' of review investigations of each and the reasoning for it. 
    $\rho$ is the median network matched-filter \ac{SNR} determined in the last completed analysis of the signal. }
    \label{tab:problem}
    \begin{tabular}{llrlp{8.2cm}}
        \toprule
        Candidate & Detectors & $\rho$ & Outcome & Reasoning \\
        \midrule
        GW230904\_152545 & H1, L1 & 0.8 & Noise & $\ln\mathcal{B} = -0.08$; posteriors are prior-dominated; no transient at the candidate time, only an elevated noise floor in both detectors. \\
        GW231013\_135504 & H1, L1 & 0.8 & Noise & Network \ac{SNR} below one, with a posterior which is prior-dominated. \\
        \addlinespace
        GW230702\_162025 & H1, L1 & 9.7 & Glitch (suspected) & Chirp mass rails at each widened bound (33, 80, then 95\,\ensuremath{M_\odot}) with $q$ at the 0.05 floor; \ac{SNR} is dominated by L1 ($\approx 9$, against $\approx 4$ in H1), where a compact 20--60\,Hz transient with no upward chirp is seen. \\
        GW231026\_130704 & H1, L1 & 7.5 & Glitch & Chirp mass rails at the lower bound and is multimodal; multiple discrete 20--100\,Hz transients in H1 at the candidate time. \\
        GW231204\_090648 & H1, L1 & 7.5 & Glitch & Periodic comb of $\sim$25--50\,Hz transients in H1 every $\sim$0.15\,s from $-8$ to $+4$\,s; L1 is clean. \\
        GW231206\_010629 & H1, L1 & 6.9 & Glitch & Compact 20--40\,Hz transient in H1 at $-0.15$\,s and 0\,s, with no counterpart in L1. \\
        GW240105\_151143 & H1 & 32.3 & Glitch & Implausibly narrow posterior ($q$ at the 0.05 bound, $d_L \approx 70$\,Mpc) from data containing repeated loud low-frequency transients. \\
        \addlinespace
        GW230728\_083628 & H1, L1 & 13.5 & Chirp mass unconstrained & Chirp mass rails against each tested prior. Strong broadband L1 transient at 20--80\,Hz and a weaker one in H1; for a signal of this total mass a real merger would also be a short burst, so glitch and signal cannot be separated by morphology. \\
        GW230824\_135331 & H1, L1 & 9.5 & Chirp mass unconstrained & Chirp mass rails against each tested prior. Compact 20--50\,Hz transient in L1, and repeated bursts of low-frequency noise in H1, suggesting an H1 glitch. \\
        GW230925\_143957 & H1, L1 & 6.8 & Chirp mass unconstrained & Chirp mass rails against each tested prior, with medians of 31, 80, and 108\,\ensuremath{M_\odot} for maxima of 33, 100, and 160\,\ensuremath{M_\odot}; the waveform fails above about 175\,\ensuremath{M_\odot}, so the bound cannot be widened further. No loud transient at the candidate time, and $\ln\mathcal{B} = 10.1$. \\
        GW231002\_143916 & H1, L1 & 8.1 & Chirp mass unconstrained & Chirp mass rails against each tested prior, with a median which moves from 79 to 101\,\ensuremath{M_\odot} as the maximum is raised from 100 to 160\,\ensuremath{M_\odot}, where it is bimodal (modes at 79 and 128\,\ensuremath{M_\odot}). No loud transient at the candidate time, and $\ln\mathcal{B} = 17.1$. \\
        GW231102\_052214 & H1, L1 & 5.6 & Chirp mass unconstrained & Chirp mass railed against each tested prior, however with a bound of 100\,\ensuremath{M_\odot} it peaks inside the prior (28--84\,\ensuremath{M_\odot}, 90\% HPD), but still rails at the upper bound, the distance rails at its upper bound, and the mass ratio is bimodal ($q \approx 0.13$ and 0.46). Network \ac{SNR} is marginal, with $\ln\mathcal{B} = 3.7$. \\
        \addlinespace
        GW230717\_102139 & H1, L1 & 7.5 & Multimodal & Chirp-mass posterior has two modes (at $\sim$25 and $\sim$30\,\ensuremath{M_\odot}) which persist when the H1 minimum frequency is raised to 32\,Hz. \\
        \bottomrule
    \end{tabular}
\end{table*}

\begin{table*}
    \centering
    \caption{The automated checks which flagged each of the candidate without ``approved'' samples, applied to the samples created by last completed analysis performed on each candidate. 
    \tfail{} indicates a failure (e.g. where the posterior rails against a prior bound), 
    \twarn{} is a warning raised for an analyst to examine, 
    and \tpass{} is a pass (i.e. there is no indication of this problem in the generated samples). 
    The remaining checks (consistency of the derived parameters, likelihood, network \ac{SNR}, and sky area) passed for every analysis in the table and for every approved analysis. 
    The last rows count the candidates flagged by each check among these candidates and among the candidates which have an approved analysis,  showing the rate at which each check flags candidates which were accepted.}
    \label{tab:matrix}
    \begin{tabular}{llccccccc}
\toprule
Event & Outcome & \shortstack{Chirp mass\\railing} & \shortstack{Distance\\railing} & \shortstack{Multiple\\modes} & \shortstack{Spins vs\\prior} & \shortstack{Noise-\\like} & \shortstack{Too\\loud} & \shortstack{One\\detector} \\
\midrule
GW230904\_152545 & Noise & \tfail & \tfail & \tpass & \tpass & \twarn & \tpass & \tpass \\
GW231013\_135504 & Noise & \tfail & \tfail & \tpass & \tpass & \twarn & \tpass & \tpass \\
\addlinespace
GW230702\_162025 & Glitch & \tfail & \tpass & \tpass & \tpass & \tpass & \tpass & \twarn \\
GW231026\_130704 & Glitch & \tfail & \tpass & \tpass & \tpass & \tpass & \tpass & \tpass \\
GW231204\_090648 & Glitch & \tfail & \tpass & \tpass & \twarn & \tpass & \tpass & \twarn \\
GW231206\_010629 & Glitch & \tfail & \tpass & \tpass & \twarn & \tpass & \tpass & \tpass \\
GW240105\_151143 & Glitch & \tpass & \tpass & \tpass & \twarn & \tpass & \twarn & \tpass \\
\addlinespace
GW230728\_083628 & Unconstrained & \tfail & \tpass & \tpass & \tpass & \tpass & \tpass & \tpass \\
GW230824\_135331 & Unconstrained & \tfail & \tpass & \tpass & \tpass & \tpass & \tpass & \tpass \\
GW230925\_143957 & Unconstrained & \tfail & \tpass & \tpass & \tpass & \tpass & \tpass & \tpass \\
GW231002\_143916 & Unconstrained & \tfail & \tpass & \twarn & \twarn & \tpass & \tpass & \tpass \\
GW231102\_052214 & Unconstrained & \tfail & \tfail & \twarn & \twarn & \tpass & \tpass & \tpass \\
\addlinespace
GW230717\_102139 & Multimodal & \tpass & \tpass & \twarn & \tpass & \tpass & \tpass & \tpass \\
\midrule
Flagged in the 13 analysed events & & 11 & 3 & 3 & 5 & 2 & 1 & 2 \\
Flagged in the 29 approved events & & 3 & 2 & 0 & 4 & 0 & 0 & 0 \\
Flagged by at least one test & & \multicolumn{7}{c}{13 of 13 analysed events; 7 of 29 approved events} \\
\bottomrule
\end{tabular}

\end{table*}

\begin{figure*}
    \centering
    \includegraphics[width=\linewidth]{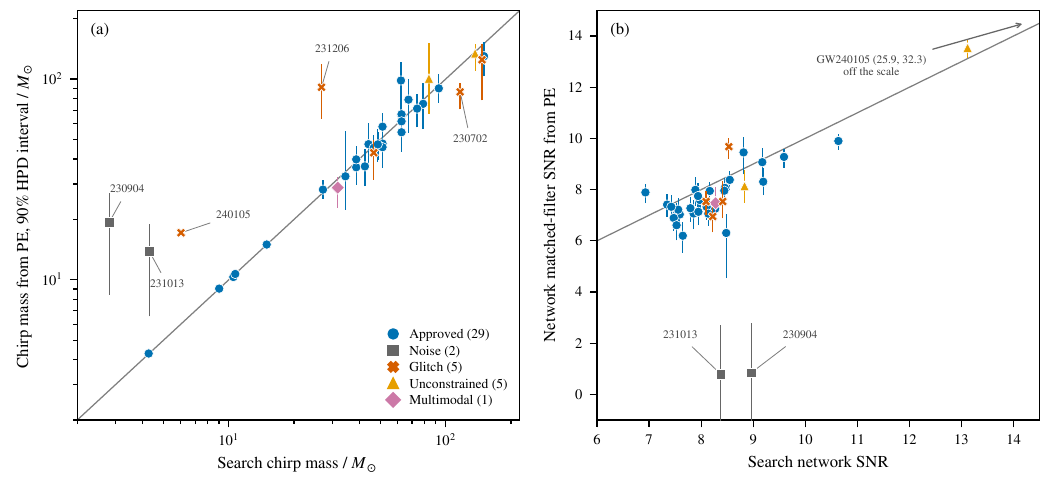}
    \caption{The chirp mass (left panel) and network \ac{SNR} (right panel) as determined by \ac{PE}, for each candidate, compared to the values found by the search. 
    For each candidate the median value is represented as a point with error bars showing the 90\% highest-posterior-density (HPD) intervals of its approved analysis, or the last analysis performed for candidates which were ultimately rejected. 
    The line shows the case where the \ac{PE} and search quantities are equal.
    Candidates are coloured by their outcome, as in Figure~\ref{fig:selection}. 
    In the right plot, GW240105\_151143, which has a search \ac{SNR} of 25.9 and a \ac{PE}-determined \ac{SNR} of 32.3, is not plotted. 
    The search values are those of the template-based pipeline which has the lowest false-alarm rate, except where that pipeline has a template at the edge of its template bank (MBTA for GW230702\_162025 and GW231002\_143916), which is not a measurement of the chirp mass, in which case the values of \texttt{gstlal} are used (see text).
    GW230824\_135331 was found only by the unmodelled cWB pipeline, which has no template chirp mass, and GW230902\_122814, GW230925\_143957, and GW231102\_052214 were found by MBTA with a template at the edge of its bank and by cWB, so they have no reliable search value and are not shown.}
    \label{fig:searchpe}
\end{figure*}

The search provides a check on the parameter estimation which is independent of our review process, as we should expect a candidate which the search found with a chirp mass and \ac{SNR} should be recovered with broadly comparable values.
Figure~\ref{fig:searchpe} shows that this is the case for most of the approved candidates.
The search chirp mass lies within the 90\% HPD interval of the distribution from \ac{PE} for 27 of the 28 approved candidates which have a reliable search value, and the median ratio of the recovered to the search \ac{SNR} is 0.94, but the chirp masses are consistent for only 4 of the 10 candidates which were not approved and have a reliable search value.
The one approved candidate which lies outwith the interval is GW230721\_222634, which lies just outside it.
The two candidates which are consistent with noise are each recovered with an \ac{SNR} of 0.8 where the search found 8.4 and 9.0, while GW240105\_151143 is recovered with a chirp mass of 17\,\ensuremath{M_\odot} and a larger \ac{SNR} where the search found 6\,\ensuremath{M_\odot}.
GW231102\_052214 was found by MBTA with a chirp mass of 11\,\ensuremath{M_\odot}, and was recovered at 28--84\,\ensuremath{M_\odot} by \ac{PE}, but the template which MBTA recovered for it lies at the edge of its template bank, so that this value is not a reliable guide (see below).
By contrast, the high-mass candidates GW230728\_083628 and GW231204\_090648 were recovered with high chirp masses comparable to those which the search found (137 and 147\,\ensuremath{M_\odot}). These lie within the parameter space of the \ac{O4a} template banks, which extend to detector-frame total masses of 610\,\ensuremath{M_\odot} for \texttt{gstlal} (which found both) and 500\,\ensuremath{M_\odot} for \texttt{pycbc} and \texttt{MBTA}~\citep{gwtc4_methods_2026}, corresponding to chirp masses of up to about 265 and 218\,\ensuremath{M_\odot} for equal masses, so that for these the railing could indicate genuine support at high mass and not a failure to recover the signal; GW230824\_135331, which was found only by the unmodelled cWB pipeline, has a rough chirp-mass estimate of 114\,\ensuremath{M_\odot} from it which is also within the \ac{PE} interval.

The search chirp mass of several of our candidates is not a reliable guide, because of the templates which MBTA recovered for them.
Five of the candidates (GW230702\_162025, GW230902\_122814, GW230925\_143957, GW231002\_143916, and GW231102\_052214) are listed in the search summary with the values of MBTA, and for each of them the MBTA template lies at the edge of its template bank, with a primary mass of 114--191\,\ensuremath{M_\odot} and a secondary mass of 2.3--4.1\,\ensuremath{M_\odot}, and so a chirp mass of only 11--19\,\ensuremath{M_\odot}.
A sixth candidate, GW230822\_230337, was also recovered by MBTA with such a template, but is listed with the values of \texttt{gstlal}.
Three of these candidates (GW230702\_162025, GW230925\_143957, and GW231002\_143916) share a single template, with masses of 191.1 and 4.1\,\ensuremath{M_\odot}, which is the template of only three of the MBTA candidates in the whole search summary, and templates with a primary mass above 100\,\ensuremath{M_\odot} and a secondary mass below 6\,\ensuremath{M_\odot} account for 97 of its 576 MBTA candidates.
Where another pipeline also found these candidates, its chirp mass is much higher: \texttt{gstlal} finds 83.8\,\ensuremath{M_\odot} for GW231002\_143916, 116.6\,\ensuremath{M_\odot} for GW230702\_162025, and 73.9\,\ensuremath{M_\odot} for GW230822\_230337, and the unmodelled cWB finds rough estimates of 62 and 52\,\ensuremath{M_\odot} for GW230925\_143957 and GW230902\_122814.
These are comparable to the chirp masses which we recover (for example a median of 71\,\ensuremath{M_\odot} for GW230822\_230337, and 101\,\ensuremath{M_\odot}, with a 90\% HPD interval of 67--151\,\ensuremath{M_\odot}, for GW231002\_143916 in an analysis with a prior maximum of 160\,\ensuremath{M_\odot} which still rails).
The search chirp mass of these candidates should therefore not be taken to indicate a low-mass system, and the chirp-mass bounds which we first set from it were too low for them.

\subsection{Candidates consistent with noise}

\begin{figure}
    \centering
    \includegraphics[width=\columnwidth]{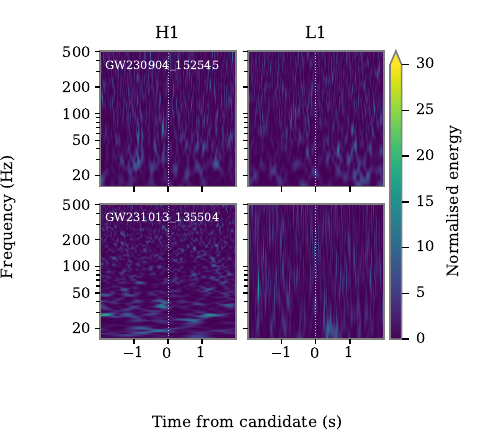}
    \caption{Whitened Q-transforms of the strain in each detector within 2\,s of the candidate time (dotted line) for the two candidates consistent with noise.}
    \label{fig:qscans-noise}
\end{figure}

GW230904\_152545 and GW231013\_135504 were found by the \texttt{pycbc} search pipeline~\citep{gwtc4_methods_2026}, with low chirp mass templates (2.8 and 4.3\,\ensuremath{M_\odot}), network \acp{SNR} of 9.0 and 8.4, and \acp{far} of 18\,yr$^{-1}$ and 64\,yr$^{-1}$.
The latter has the highest \ac{far} of any of our candidates.
We find that both have a network matched-filter \ac{SNR} of 0.8 in our \ac{PE} analyses (with a 90\% HPD interval from $-1.1$ to 2.8), and a signal-to-noise Bayes factor of $\ln\mathcal{B} = -0.08$ and $-0.04$, indicating that the data slightly favour the noise hypothesis.
Their posteriors are indistinguishable from the prior, with a Kullback--Leibler divergence (see Section~\ref{sec:information}) of at most 0.03\,bits for every parameter, and 0.8 and 0.6\,bits for all of the parameters together, and the Q-transforms (shown in Figure~\ref{fig:qscans-noise}) do not obviously show a transient at the candidate time in either detector.
We conclude that these candidates are consistent with noise, although their values of $\pastro$ are 0.74 and 0.54, and we do not report parameters for them.

\subsection{Glitch-affected candidates}

\begin{figure}
    \centering
    \includegraphics[width=\columnwidth]{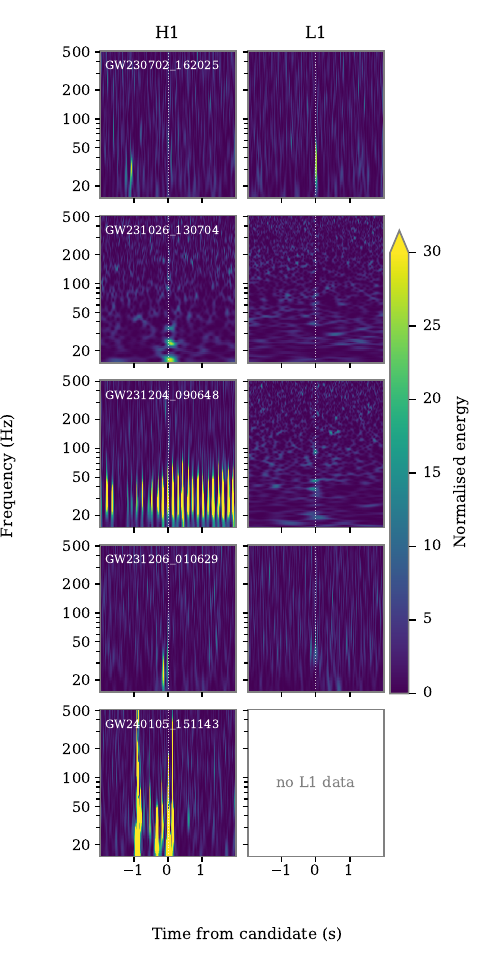}
    \caption{Whitened Q-transforms of the strain in each detector within 2\,s of the candidate time (dotted line) for the glitch-affected candidates. GW240105\_151143 has no L1 data. The colour scale saturates at a normalised energy of 30.}
    \label{fig:qscans-glitch}
\end{figure}

Five candidates show evidence that the candidate is dominated by a non-Gaussian noise transient (Figure~\ref{fig:qscans-glitch} and Table~\ref{tab:problem}).
GW231204\_090648 has a periodic comb of transients at about 25--50\,Hz, every 0.15\,s or so, which extends from several seconds before to several seconds after the candidate in H1, while L1 is clean. 
The noise level of H1 below 50\,Hz is around six times greater than that in L1, which is the band in which contains most of this signal.
Thus H1 contributes little to the \ac{SNR} and L1 contributes 89\% of the squared \ac{SNR} (2.4 in H1 and 6.7 in L1) of the candidate.
We did not perform a single-detector analysis on L1 to determine if this provides an acceptable posterior.

GW231206\_010629 has a compact transient at 20--40\,Hz in H1, shortly before and at the candidate time, with no counterpart in L1, and GW231026\_130704 has several discrete transients at 20--100\,Hz in H1 at the candidate time; its first analysis rails at the lower bound of the chirp mass and is multimodal.
The \ac{SNR} of GW230702\_162025 is dominated by L1, in which there is a compact transient at 20--60\,Hz without an obvious chirp morphology, and its chirp mass rails against each prior distribution we tested (with upper limits of 33, 80, and 95\,\ensuremath{M_\odot}) with the mass ratio railing at its lower bound of 0.05.
This could be a very massive merger of which only the merger and ringdown are in band.
Such a signal has been confidently observed: GW231123, with a total mass of 190--265\,\ensuremath{M_\odot}, has only about five cycles between 30 and 80\,Hz~\citep{gw231123}.
However, GW231123 was detected coherently in both LIGO detectors, and its authors note that without this coincidence it would likely have been dismissed as a glitch; its analysis also needed H1 data from which a glitch had been subtracted, and a comparison of several waveform models.
In GW230702\_162025 by contrast the \ac{SNR} is concentrated in a single detector, and we cannot show that the transient is a signal and not a glitch.
GW240105\_151143 was found in H1 only, with a search \ac{SNR} of 25.9, and we determine the median \ac{SNR} from \ac{PE} as 32.3.
Its posterior on the mass ratio rails at the prior bound, while \ac{PE} implies a very small luminosity distance: around 70\,Mpc.
The signal-to-noise Bayes factor is $\ln\mathcal{B} = 452$, and the divergence of the prior from the posterior is between 5 and 42\,bits for each parameter, and 79\,bits for all of the parameters together, the most of any of our candidates.
However, inspection of the Q-scan of the data show they contain repeated loud transients at low frequency.
For three of these (GW230702\_162025, GW231206\_010629, and GW240105\_151143) the recovered chirp mass is inconsistent with that of the search (see Figure~\ref{fig:searchpe}), and we rejected all five.
The Q-scans for each of these candidates are shown in Figure~\ref{fig:qscans-glitch}.

\subsection{Candidates where the chirp mass is not constrained}

For five candidates the chirp mass posterior samples rail against the upper bound of the prior distribution.
When we increased the upper bound on the prior this behaviour persisted, implying that the data were not informing the inference of this parameter.

For GW230728\_083628 and GW230824\_135331 our \ac{PE} analyses infer that each has a detector-frame total mass of several hundred solar masses (with median values of 490 and 520\,\ensuremath{M_\odot} respectively).
The frequency of the corresponding innermost stable circular orbit is below the 20\,Hz minimum frequency used in constructing the likelihood for essentially every sample in these posteriors.
Therefore only the merger and ringdown would be in band, and so the mass ratios, with medians of 0.16 and 0.15, and the chirp masses are poorly constrained.

We found that raising the upper chirp mass prior bound for GW230824\_135331 from 128\,\ensuremath{M_\odot} to 160\,\ensuremath{M_\odot} moved the median from 117 to 144\,\ensuremath{M_\odot}, with the posterior railing against the upper bound in both cases.
This candidate was found only by the unmodelled \texttt{cWB} pipeline, so we have no template value to compare to.

We made the largest number of attempts, jointly with GW230830\_064744, to produce a posterior distribution for GW231102\_052214.
We tested chirp mass prior maxima of 25, 29.8, 60, and 71.5\,\ensuremath{M_\odot}, which resulted in the posterior median chirp masses of 23, 27, 44, and 55\,\ensuremath{M_\odot} respectively, with each set of posterior samples railing against the upper bound.
With a maximum of 100\,\ensuremath{M_\odot} the posterior peaks inside the prior, with a median of 61\,\ensuremath{M_\odot} and a 90\% HPD interval of 28--84\,\ensuremath{M_\odot}, but still rails at the upper bound.
However, its distance rails at the upper bound, and its mass ratio is bimodal ($q \approx 0.13$ and 0.46).
The only template-based pipeline which found it, MBTA, recovered a chirp mass of 11\,\ensuremath{M_\odot} from a template at the edge of its bank, which lies below the posterior but which is not a reliable value (see the discussion of Figure~\ref{fig:searchpe}), and its network \ac{SNR} of 5.6 is the lowest of any candidate which we did not otherwise find to be consistent with noise.

The chirp masses of GW230925\_143957 and GW231002\_143916 followed the upper bound in the same way, after our first priors proved too low because the MBTA templates of these candidates lie at the edge of its template bank (see the discussion of Figure~\ref{fig:searchpe}).
For GW230925\_143957 the posterior median was 31, 80, and 108\,\ensuremath{M_\odot} for prior maxima of 33, 100, and 160\,\ensuremath{M_\odot}, with the posterior railing against the upper bound for each.
For GW231002\_143916 the median was 79\,\ensuremath{M_\odot} for a maximum of 100\,\ensuremath{M_\odot} and 101\,\ensuremath{M_\odot} for a maximum of 160\,\ensuremath{M_\odot}, for which the posterior has two modes of similar weight, at 79 and 128\,\ensuremath{M_\odot}, the lower of which is consistent with the chirp mass of 83.8\,\ensuremath{M_\odot} of the \texttt{gstlal} template.
We could not raise the bound further, because the waveform fails for heavier systems (Section~\ref{sec:config}).
The network \acp{SNR} of these two candidates are 6.8 and 8.1, and their signal-to-noise Bayes factors $\ln\mathcal{B}$ are 10.1 and 17.1, so that the data contain more than noise, but our analyses do not constrain the chirp mass.
Their Q-transforms show no transient as loud as those of GW230728\_083628 and GW230824\_135331, so we cannot attribute the behaviour to a glitch; we do not report posterior samples for them.

\begin{figure}
    \centering
    \includegraphics[width=\columnwidth]{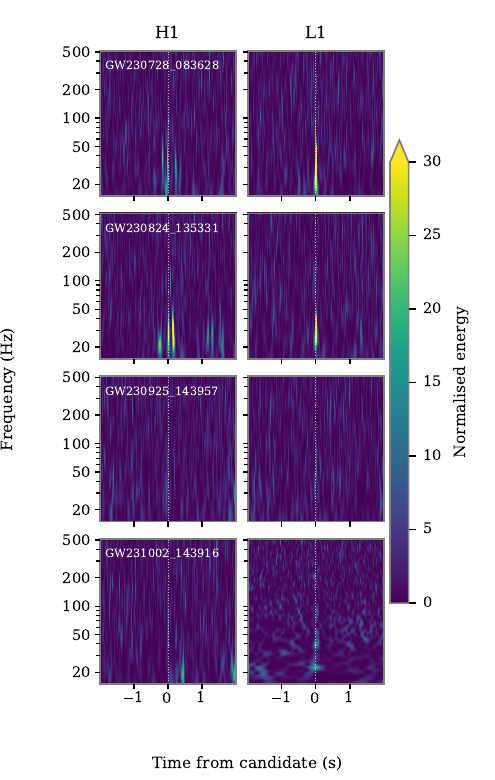}
    \caption{Whitened Q-transforms of the strain in each detector within 2\,s of the candidate time (dotted line) for four of the candidates where the chirp mass is not constrained.}
    \label{fig:qscans-unconstrained}
\end{figure}

We show the Q-scans for four of these candidates in Figure~\ref{fig:qscans-unconstrained}.

\subsection{Candidate with multimodal posteriors}

GW230717\_102139 has a network \ac{SNR} of 7.5.
The chirp-mass posterior samples of its first analysis has a main mode at 30\,\ensuremath{M_\odot} and a weaker second mode at 25.5\,\ensuremath{M_\odot} which holds about 13\% of the probability and is present in all of the parallel chains.
Its arrival-time posterior is unimodal, so that both modes correspond to the same event time.
We rejected the initial analysis of this candidate because of the multimodality, and further investigation identified excess noise between 20 and 70\,Hz in H1.
We performed a second analysis in which we raised the minimum frequency of H1 to 32\,Hz.
However, in this second analysis the second mode persists and is stronger, with modes at 24.8 and 30.0\,\ensuremath{M_\odot} which hold 39\% and 61\% of the probability.
The lower mode has a median mass ratio of 0.43, an effective spin of $-0.14$, and a distance of 2.2\,Gpc, against 0.68, $0.13$, and 3.3\,Gpc for the upper mode, and the merger frequencies of the two are about 70 and 60\,Hz respectively.
As a result of the continued presence of multimodality in the posterior we rejected this candidate.
Rejecting this candidate due to multimodality is a purposefully conservative choice; we note that signals which have been broadly accepted within the community show similar behaviour, such as GW231123~\citep{gw231123}.

%%%%%%% Daniel has reviewed up to here 2026-10-20

\section{Source parameters}\label{sec:sourceparameters}

Having completed the various checks required for us to conclude that the posterior distributions produced by the analyses described in section~\ref{sec:pe} are viable we are able to report the likely properties of the sources of the signals we have analysed, assuming that they are in fact of astrophysical origin.
Unless stated otherwise we state values for these properties as the median \emph{a posteriori} value, with  credible intervals representing 90\% highest-posterior-density intervals.

\begin{table*}
    \centering
    \footnotesize
    \caption{The median and the 90\% HPD credible interval of the source-frame chirp mass, mass ratio, effective and precessing spin, and luminosity distance of the events with an approved analysis, with the median network matched-filter \ac{SNR} $\rho$ and the information gained about all of the sampled parameters, $D_\mathrm{KL}$ (Section~\ref{sec:information}).}
    \label{tab:summary}
    \begin{tabular}{lccccccc}
\toprule
Event & $\rho$ & $\mathcal{M}_\mathrm{src}\,/\,M_\odot$ & $q$ & $\chi_\mathrm{eff}$ & $\chi_p$ & $d_L\,/\,\mathrm{Mpc}$ & $D_\mathrm{KL}\,/\,\mathrm{bit}$ \\
\midrule
GW230531\_141100 & 7.6 & $34.9^{+8.4}_{-7.7}$ & $0.71^{+0.29}_{-0.26}$ & $0.09^{+0.29}_{-0.26}$ & $0.47^{+0.40}_{-0.35}$ & $3915^{+2834}_{-2302}$ & 16.5 \\
GW230603\_174756 & 7.1 & $17.9^{+3.6}_{-3.4}$ & $0.74^{+0.26}_{-0.27}$ & $0.06^{+0.24}_{-0.23}$ & $0.46^{+0.39}_{-0.36}$ & $3376^{+2216}_{-1950}$ & 20.7 \\
GW230606\_024545 & 7.0 & $22.0^{+4.9}_{-4.2}$ & $0.64^{+0.34}_{-0.26}$ & $-0.09^{+0.28}_{-0.33}$ & $0.39^{+0.37}_{-0.34}$ & $3913^{+2507}_{-2247}$ & 18.7 \\
GW230609\_010824 & 8.0 & $29.3^{+7.5}_{-6.4}$ & $0.46^{+0.32}_{-0.23}$ & $0.09^{+0.27}_{-0.24}$ & $0.28^{+0.33}_{-0.25}$ & $3590^{+3098}_{-2179}$ & 20.2 \\
GW230615\_160825 & 7.2 & $63.2^{+22.0}_{-17.6}$ & $0.70^{+0.30}_{-0.32}$ & $0.11^{+0.42}_{-0.36}$ & $0.47^{+0.37}_{-0.37}$ & $6908^{+4823}_{-4482}$ & 15.2 \\
GW230618\_102549 & 7.0 & $38.4^{+13.4}_{-11.3}$ & $0.65^{+0.35}_{-0.30}$ & $0.04^{+0.37}_{-0.33}$ & $0.48^{+0.39}_{-0.39}$ & $6167^{+4408}_{-3890}$ & 15.8 \\
GW230624\_214944 & 9.9 & $12.6^{+1.0}_{-1.0}$ & $0.33^{+0.33}_{-0.12}$ & $-0.07^{+0.17}_{-0.26}$ & $0.25^{+0.38}_{-0.22}$ & $981^{+451}_{-485}$ & 24.0 \\
GW230625\_211655 & 7.1 & $27.2^{+6.8}_{-5.9}$ & $0.70^{+0.30}_{-0.28}$ & $0.07^{+0.31}_{-0.30}$ & $0.48^{+0.40}_{-0.37}$ & $4518^{+2905}_{-2687}$ & 20.7 \\
GW230708\_071859 & 7.3 & $25.7^{+6.4}_{-5.4}$ & $0.76^{+0.24}_{-0.25}$ & $-0.07^{+0.26}_{-0.30}$ & $0.50^{+0.39}_{-0.37}$ & $4836^{+2940}_{-2863}$ & 18.7 \\
GW230709\_063445 & 6.9 & $32.3^{+9.1}_{-8.4}$ & $0.71^{+0.29}_{-0.27}$ & $0.01^{+0.31}_{-0.31}$ & $0.46^{+0.39}_{-0.36}$ & $5781^{+3943}_{-3729}$ & 16.4 \\
GW230721\_222634 & 7.2 & $50.1^{+17.2}_{-14.1}$ & $0.43^{+0.35}_{-0.29}$ & $0.05^{+0.39}_{-0.36}$ & $0.34^{+0.38}_{-0.30}$ & $5736^{+4446}_{-3970}$ & 16.0 \\
GW230723\_084820 & 7.9 & $36.0^{+10.0}_{-8.4}$ & $0.69^{+0.31}_{-0.28}$ & $0.00^{+0.28}_{-0.29}$ & $0.43^{+0.38}_{-0.36}$ & $5022^{+3364}_{-2992}$ & 19.3 \\
GW230807\_205045 & 7.1 & $47.8^{+12.3}_{-11.2}$ & $0.74^{+0.26}_{-0.31}$ & $0.03^{+0.41}_{-0.35}$ & $0.48^{+0.37}_{-0.37}$ & $5527^{+3577}_{-3128}$ & 15.5 \\
GW230817\_212349 & 7.3 & $24.0^{+5.3}_{-5.3}$ & $0.61^{+0.36}_{-0.26}$ & $0.09^{+0.30}_{-0.27}$ & $0.42^{+0.38}_{-0.34}$ & $3974^{+2765}_{-2272}$ & 19.0 \\
GW230822\_230337 & 7.9 & $39.1^{+10.1}_{-8.8}$ & $0.72^{+0.28}_{-0.28}$ & $0.07^{+0.32}_{-0.32}$ & $0.51^{+0.38}_{-0.38}$ & $5152^{+3141}_{-2853}$ & 17.6 \\
GW230823\_142524 & 9.5 & $29.5^{+6.5}_{-6.5}$ & $0.41^{+0.37}_{-0.25}$ & $-0.41^{+0.38}_{-0.31}$ & $0.54^{+0.38}_{-0.28}$ & $1172^{+1262}_{-616}$ & 29.2 \\
GW230830\_064744 & 6.3 & $20.0^{+8.4}_{-6.1}$ & $0.48^{+0.48}_{-0.28}$ & $0.28^{+0.37}_{-0.35}$ & $0.46^{+0.36}_{-0.36}$ & $3869^{+3968}_{-2593}$ & 18.7 \\
GW230831\_134621 & 8.4 & $8.0^{+0.8}_{-0.6}$ & $0.65^{+0.34}_{-0.26}$ & $-0.10^{+0.16}_{-0.15}$ & $0.32^{+0.34}_{-0.27}$ & $1557^{+649}_{-705}$ & 28.2 \\
GW230902\_122814 & 6.6 & $51.7^{+23.4}_{-17.2}$ & $0.69^{+0.31}_{-0.31}$ & $0.06^{+0.39}_{-0.33}$ & $0.46^{+0.39}_{-0.36}$ & $7395^{+6450}_{-5848}$ & 14.8 \\
GW230902\_172430 & 8.1 & $8.0^{+0.9}_{-0.7}$ & $0.64^{+0.33}_{-0.26}$ & $0.15^{+0.13}_{-0.10}$ & $0.38^{+0.36}_{-0.29}$ & $1824^{+851}_{-824}$ & 29.6 \\
GW230902\_224555 & 6.6 & $34.6^{+11.2}_{-10.5}$ & $0.61^{+0.39}_{-0.29}$ & $0.07^{+0.35}_{-0.31}$ & $0.43^{+0.40}_{-0.36}$ & $5791^{+4598}_{-3870}$ & 15.9 \\
GW230920\_064709 & 9.1 & $3.9^{+0.2}_{-0.2}$ & $0.24^{+0.13}_{-0.13}$ & $0.08^{+0.26}_{-0.20}$ & $0.16^{+0.18}_{-0.14}$ & $515^{+311}_{-216}$ & 42.2 \\
GW231005\_144455 & 7.4 & $27.8^{+7.0}_{-5.7}$ & $0.71^{+0.29}_{-0.25}$ & $0.01^{+0.29}_{-0.29}$ & $0.46^{+0.40}_{-0.36}$ & $4431^{+3057}_{-2471}$ & 18.6 \\
GW231102\_232433 & 6.2 & $37.2^{+13.8}_{-12.9}$ & $0.56^{+0.38}_{-0.29}$ & $0.13^{+0.40}_{-0.32}$ & $0.42^{+0.39}_{-0.35}$ & $7229^{+6187}_{-5214}$ & 15.9 \\
GW231113\_150041 & 7.7 & $33.4^{+10.5}_{-9.2}$ & $0.51^{+0.40}_{-0.29}$ & $-0.00^{+0.30}_{-0.33}$ & $0.56^{+0.39}_{-0.39}$ & $5146^{+3611}_{-3050}$ & 16.9 \\
GW231120\_022103 & 9.3 & $7.3^{+0.6}_{-0.5}$ & $0.72^{+0.28}_{-0.26}$ & $0.03^{+0.11}_{-0.07}$ & $0.30^{+0.32}_{-0.27}$ & $1221^{+488}_{-549}$ & 31.0 \\
GW231126\_010928 & 8.0 & $28.4^{+6.4}_{-5.8}$ & $0.74^{+0.26}_{-0.27}$ & $0.11^{+0.27}_{-0.25}$ & $0.47^{+0.39}_{-0.37}$ & $3894^{+2345}_{-2199}$ & 20.3 \\
GW231220\_173406 & 7.3 & $29.1^{+8.1}_{-7.1}$ & $0.63^{+0.34}_{-0.25}$ & $0.04^{+0.28}_{-0.29}$ & $0.41^{+0.40}_{-0.35}$ & $5442^{+4203}_{-3549}$ & 18.1 \\
GW231231\_120147 & 8.3 & $27.6^{+7.1}_{-6.2}$ & $0.63^{+0.35}_{-0.28}$ & $-0.04^{+0.30}_{-0.33}$ & $0.65^{+0.33}_{-0.38}$ & $3796^{+2908}_{-2278}$ & 20.4 \\
\bottomrule
\end{tabular}

\end{table*}

\begin{figure*}
    \centering
    \includegraphics[width=\linewidth]{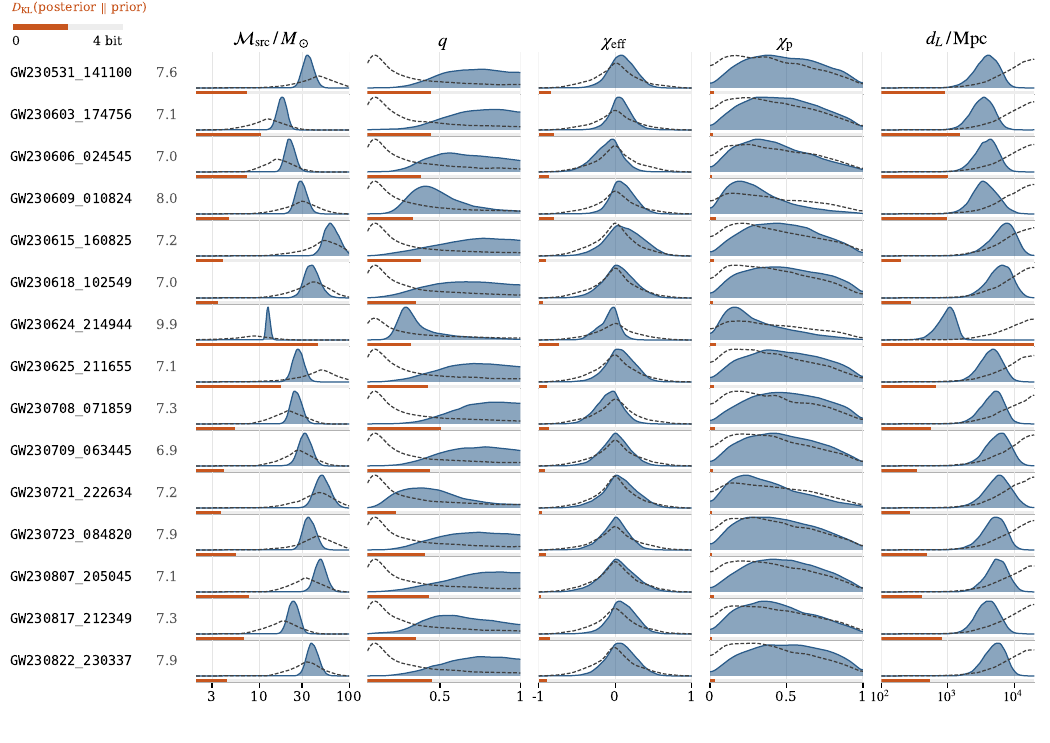}
    \caption{The marginal posterior distributions (filled) of the source-frame chirp mass $\mathcal{M}_\mathrm{src}$, mass ratio $q$, effective spin $\chi_\mathrm{eff}$, precessing spin $\chi_\mathrm{p}$, and luminosity distance $d_L$ for each candidate with an approved analysis, together with the prior used in the analysis (dashed line). In each panel the posterior and prior share one vertical scale, so a posterior which is taller than its prior is more concentrated than it. Candidates are shown in chronological order, and the median network matched-filter \ac{SNR} is given beside each name. 
    The bar beneath each panel represents the Kullback--Leibler divergence of the prior from the posterior (Section~\ref{sec:information}), in bits; the full width of the panel corresponds to four bits. The remaining candidates are shown in Figure~\ref{fig:multiples2}. The densities are kernel density estimates of the samples, in $\ln d_L$ for the distance and $\ln\mathcal{M}_\mathrm{src}$ for the chirp mass, reflected at the hard bounds of the prior (including the maximum distance of 20\,Gpc).}
    \label{fig:multiples}
\end{figure*}

\begin{figure*}
    \centering
    \includegraphics[width=\linewidth]{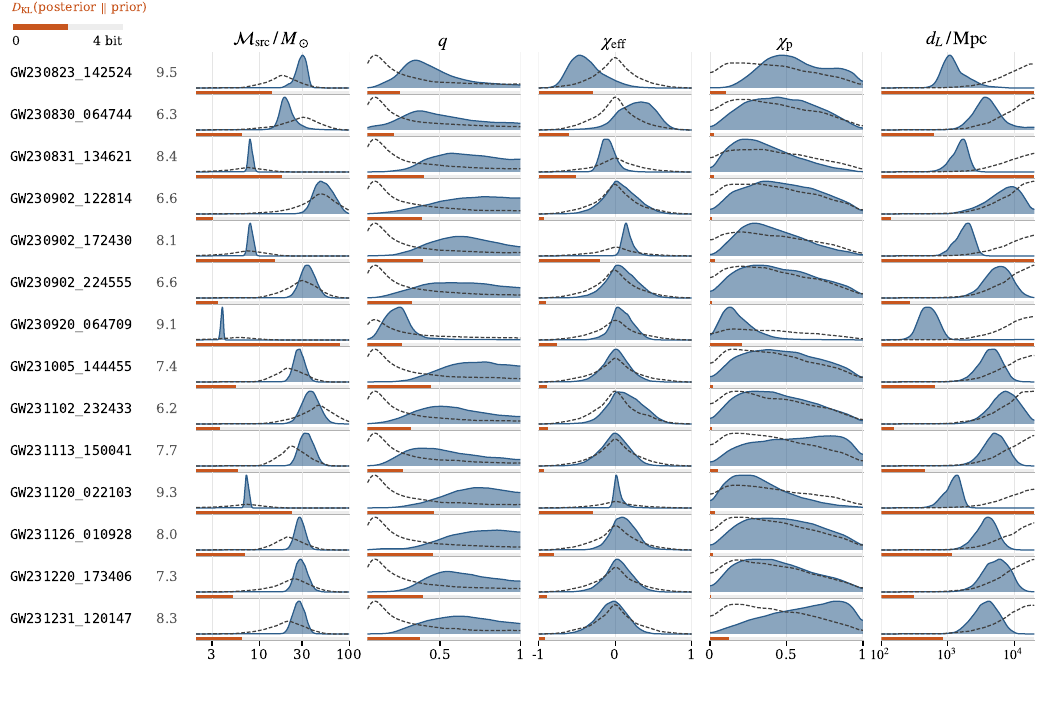}
    \caption{As Figure~\ref{fig:multiples}, for the remaining candidates.}
    \label{fig:multiples2}
\end{figure*}

Table~\ref{tab:summary} provides a summary of the properties for each of the \napproved{} candidates for which we have an approved analysis, while Figures~\ref{fig:multiples} and~\ref{fig:multiples2} show the marginal posterior distributions for chirp mass in the source frame, $\mathcal{M}_{\rm src}$, mass ratio, $q$, effective spin, $\chi_{\rm eff}$, precessing spin, $\chi_{\rm p}$, and luminosity distance, $D_{\rm L}$, together with the prior used in the analysis.

\subsection{Information gained from the data}\label{sec:information}

The nature of these signals, being of close-to marginal significance, implies that they are not likely to be strongly informative when parameter estimation is performed.
The posterior distributions are generally broad, which leads to the question of how much we actually learn from these observations.
We quantify the information gained from the data using the Kullback--Leibler divergence of the prior $\pi$ from the posterior $p$,
\begin{equation}
    \label{eq:kl-divergence}
    D_\mathrm{KL}(p \,\|\, \pi) = \int p(\theta) \log_2 \frac{p(\theta)}{\pi(\theta)} \, \mathrm{d}\theta,
\end{equation}
which is the information gained in moving from the prior to the posterior.
We take the logarithm to base two, so that it is in bits.
It is zero for a posterior equal to the prior, is not symmetric between the two, and, unlike the Jensen--Shannon divergence, it has no upper limit.
We use it in two ways.
For each parameter we evaluate it for the marginal posterior and prior, using kernel density estimates of the samples which are reflected at hard bounds (the maximum distance of the prior, and the limits of $q$, $\chi_\mathrm{eff}$, and $\chi_p$), over the full range of the samples.
We also evaluate it for the posterior over all of the sampled parameters together, which is the information $H$ reported by the nested sampler.
We calculate this from the samples as $H = \langle \ln \mathcal{L} \rangle_p - \ln\mathcal{B}$ (converted to bits), where $\mathcal{L}$ is the likelihood ratio of the signal to the noise hypothesis, the average is over the posterior samples, and $\mathcal{B}$ is the signal-to-noise Bayes factor.
The luminosity distance is marginalised over in the likelihood, so this total does not include it.

\begin{figure*}
    \centering
    \includegraphics[width=\linewidth]{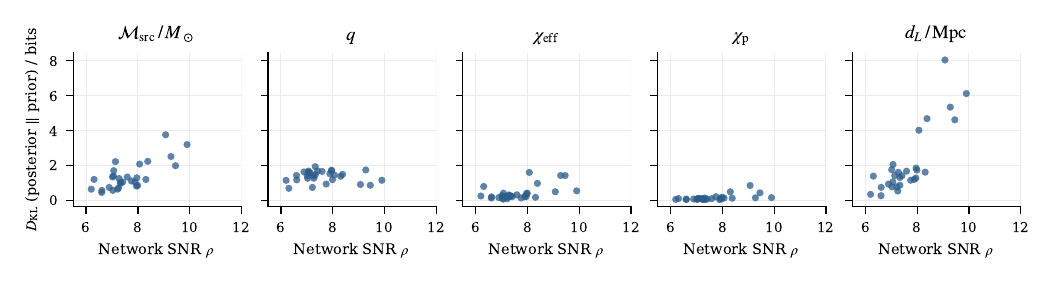}
    \caption{The Kullback--Leibler divergence of the prior from the posterior of each parameter, against the network matched-filter \ac{SNR}, for each of the candidates with an approved analysis.}
    \label{fig:divergence}
\end{figure*}

Figure~\ref{fig:divergence} shows the divergence for each candidate.
For the analysed signals the median divergences are 1.19\,bit for the chirp mass, 1.40\,bit for the distance, 1.42\,bit for the mass ratio, 0.24\,bit for $\chi_\mathrm{eff}$, and 0.09\,bit for $\chi_p$, so we generally gain very little information about the spin parameters from these analyses.
Our analyses typically exclude small mass ratios, and this results in the divergence of the mass ratio being between 0.7 and 1.9\,bit regardless of the \ac{SNR} of the signals.
The prior, which is uniform in the component masses, results in 51\% of the prior probability being below $q = 0.2$ for every candidate (between 0.49 and 0.52), whereas the posteriors place a median of 1\% there.
If we calculate the divergence from only two bins: $q$ below and above 0.2, the median divergence is 0.95\,bits, and we find the posterior probability below 0.2 does not depend strongly on the \ac{SNR}.

\begin{figure}
    \centering
    \includegraphics[width=\linewidth]{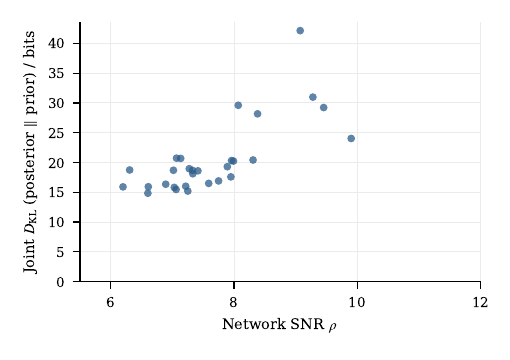}
    \caption{The information gained about all of the sampled parameters together, the Kullback--Leibler divergence of the joint prior from the joint posterior, against the network matched-filter \ac{SNR}, for each of the candidates with an approved analysis.}
    \label{fig:information}
\end{figure}

The total information gained, shown in Figure~\ref{fig:information}, has a median of 18.7\,bits (an interquartile range of 16.4--20.7\,bits) and ranges from 14.8 to 42.2\,bits.
It increases with the \ac{SNR}, with a correlation of 0.72, at roughly 5\,bits for each unit of \ac{SNR}.
The most informative candidate is GW230920\_064709, with 42.2\,bits at a network \ac{SNR} of 9.1, some 15\,bits more than the trend with \ac{SNR} would suggest.
It has the largest divergence for the chirp mass (3.8\,bits) and the distance (8.0\,bits) of any approved candidate, consistent with its low mass (see Section~\ref{sec:mass}), for which the signal spends many more cycles in the sensitive band of the detectors than it does for the other candidates.
The next most informative are GW231120\_022103 (31.0\,bits), GW230902\_172430 (29.6\,bits), GW230823\_142524 (29.2\,bits), and GW230831\_134621 (28.2\,bits); GW230823\_142524 is also the candidate with the most precession and the most support for negative $\chi_\mathrm{eff}$ (Section~\ref{sec:spin}).
The least informative are GW230902\_122814 (14.8\,bits, with a network \ac{SNR} of 6.6), GW230615\_160825 (15.2\,bits), GW230807\_205045 (15.5\,bits), GW230618\_102549 (15.8\,bits), and GW230902\_224555 and GW231102\_232433 (both 15.9\,bits), all of which have network \acp{SNR} between 6.2 and 7.2.
Even these gain nearly 15\,bits; we expect that a substantial part of this is for the extrinsic parameters, such as the time of arrival and the sky position, which a marginal signal can still localise, and not for the intrinsic parameters of the source.
The total information does not distinguish the candidates which we approved from those which we did not: the candidates which we rejected as unconstrained, multimodal, or glitches have between 14.4 and 23.8\,bits, which is within the range of the approved candidates, with the exception of GW240105\_151143 (79\,bits).
Only the two candidates consistent with noise stand apart, with 0.8 and 0.6\,bits.

\subsection{Mass parameters}\label{sec:mass}

\begin{figure*}
    \centering
    \includegraphics[width=\linewidth]{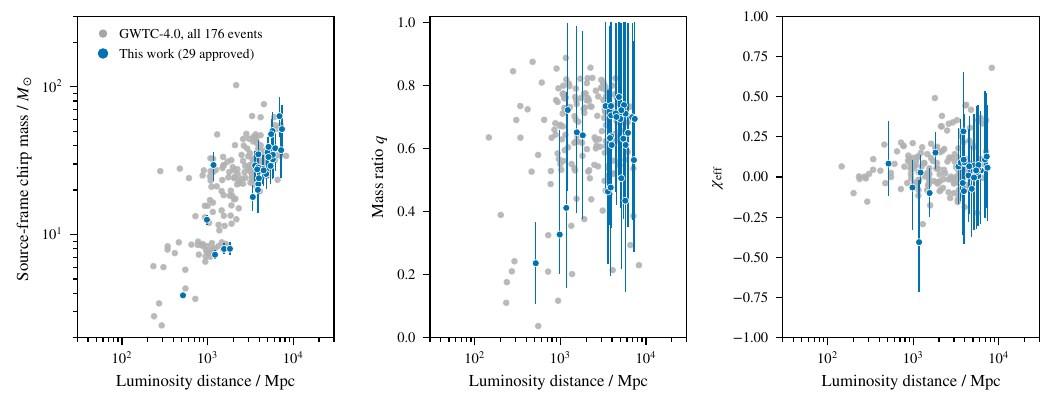}
    \caption{The source-frame chirp mass, mass ratio, and effective spin against luminosity distance for the candidates with an approved analysis (blue) and for the candidates in \ac{GWTC4} (grey), spanning the first three observing runs and the first part of the fourth. Points are medians and bars the 90\% intervals of the parameter on the vertical axis (HPD intervals for the candidates of this work, and the intervals as published for the catalogue). }
    \label{fig:context}
\end{figure*}

The parameters we infer for these candidates span a comparable range to the candidates in \ac{GWTC4}, and we show them in context with those candidates in figure~\ref{fig:context}.
The bulk of the candidates are consistent with being produced by high mass binaries.
The median total source-frame mass is 71\,\ensuremath{M_\odot}, compared to 62\,\ensuremath{M_\odot}; and the median mass ratio is 0.65, compared to 0.67 for all candidates in \ac{GWTC4} which previous had \ac{PE} results.

GW230920\_064709, however, stands out from the other candidates.
We infer that is has a source-frame chirp mass of $3.88^{+0.15}_{-0.21}\,M_\odot$, which is consistent with being an \ac{NSBH}, as we infer the secondary mass to lie in the range $1.7$--$2.8\,\ensuremath{M_\odot}$, which is broadly comparable with anticipated \ac{NS} masses.
This is consistent with the results from the search reported in~\citet{catalog-gwtc-4d0, gwtc4-search-datarelease}, where the candidate is reported as having a probability of $0.65$ of being an \ac{NSBH} and $0.16$ of being a \ac{BBH}.
We have treated this candidate as a \ac{BBH} during analysis. 
The IMRPhenomXPHM model we used does not include tidal effects, however we don't anticipate that these would be measurable in a signal at this \ac{SNR} (9.1) (see~\citealp{MeasuringNeutronStar} for a discussion of the measurability of tidal effects from parameter estimation).

\subsection{Distance}

In common with most candidates in \ac{GWTC4}, the luminosity distance is poorly measured for most of the candidates.
Measuring this parameter is complicated as a result of the degeneracy between the luminosity distance and the source's inclination angle.
For signals with high \ac{SNR} information from the higher-order modes of the signal can help to break this degeneracy, but by the nature of our selection, these candidates all have low network \acp{SNR}.

The median luminosity distance which we infer for the candidates is 4.4\,Gpc, corresponding to a redshift of $z=0.71$.
For 9 of the candidates the 95th percentile of the distance lies beyond 10\,Gpc, with the largest being 15.2\,Gpc.
This means that these candidates are amongst the most distant of those reported in \ac{GWTC4}.
This is not unexpected, since the amplitude of the signal is inversely proportional to the distance, so the most distant candidates will also generally be those with the lowest \acp{SNR}.

%%%% Daniel has reviewed up to here 2026-10-26

The posteriors of two of them, GW230902\_122814 and GW231102\_232433, extend to the upper prior bound of 20\,Gpc, with 5\% of the samples of each beyond 15\,Gpc, so that for these the posterior is partly set by the prior, as it can be for the weakest candidates of a catalogue~\citep{catalog-gwtc3}.
GW230902\_122814, whose network \ac{SNR} of 6.6 is among the lowest of the approved candidates, is the clearest case: its chirp mass is constrained (a detector-frame median of 110\,\ensuremath{M_\odot}, with a 90\% interval of 83--132\,\ensuremath{M_\odot}), which is why we approved it, but its distance of $7.4^{+6.5}_{-5.8}$\,Gpc, a redshift of $1.07^{+0.73}_{-0.75}$, and so its source-frame masses, are poorly measured.

The prior on the distance is uniform in comoving volume and source-frame time, and so it places most of its probability at distances larger than the posteriors do, as can be seen in Figure~\ref{fig:multiples}.
Signals with low-SNR are anticipated to show behaviour like this under some circumstances, and a comprehensive discussion will be given in~\citet{WilliamsVeitchConvictions}.
Similar behaviour has previously been observed in some of the lowest significance candidates for which \ac{PE} was provided in \ac{GWTC3}, but in general inference from candidates such as these should be treated with caution if being used to inform other analyses.

%%%%

\subsection{Spin parameters}\label{sec:spin}

As spin is not measured as easily from the signal as the mass information, we should anticipate that in relatively low-\ac{SNR} candidates, such as those analysed in this work, we will recover relatively little information about spin parameters.
However, some candidates do show deviation from the prior probability distribution.
GW230902\_172430 shows strong support for positive $\chi_{\mathrm{eff}}$, indicative of a binary with aligned spins: the median is 0.15, the 90\% interval is 0.05--0.28, and 99.6\% of the posterior lies above zero.
GW230823\_142524 is the candidate which shows most support for negative $\chi_{\mathrm{eff}}$, on the other hand, with a median of $-0.41$, a 90\% interval of $-0.71$ to $-0.02$, and 6.7\% of the posterior above zero.
This candidate is therefore comparable to GW241110\_124123, observed during \ac{O4b}, which similarly has very strong support for anti-aligned spins~\citep{gw241011}.
None of the other candidates has more than 90\% of its posterior on one side of zero.

The precessing spin parameter is harder still to measure from the data, and the posteriors of $\chi_{\mathrm{p}}$ are for nearly all of our candidates close to the prior.
We also quantify the evidence for precession by the precessing \ac{SNR}, $\rho_p$, of the network, for which values larger than about 2 are usually taken to show that precession can be measured~\citep{fairhurst_precession_2020, fairhurst_twoharmonic_2020}.
The median $\rho_p$ of the candidates is 0.5.
Only GW230823\_142524, the candidate which also has strong support for negative $\chi_{\mathrm{eff}}$, has a posterior which is mostly above $\rho_p=2$, with a median of 4.0 and 86\% of the posterior above 2.1.

\section{Conclusions}

In this work we have analysed gravitational-wave strain data from \nanalysed{} signal candidates which were contained within \ac{GWTC4} but where parameter estimation results were not published in that version of the catalogue, since these candidates failed to exceed a statistical significance threshold based on their \ac{far} (however they are found to be more likely to be of astrophysical origin than not, all having $\pastro\geq 0.5$).
Unsurprisingly, given the lower significance of these candidates, which also generally have lower $\pastro$ than those candidates where \ac{GWTC4} did provide parameter estimation results, we found that performing a standard analysis on some of the signals (\nnoapproved{}) did not provide a sensible posterior probability distribution.
We produced posteriors which appear to be reasonable for \napproved{} of the candidates, though we note that this does not equate to the signal being of astrophysical origin, nor does a candidate not having a posterior in this work imply that it is non-astrophysical.
We made a conscious choice not to pursue labour-intensive activities which cannot currently be automated, such as deglitching the strain data.
Doing these may have permitted comprehensive parameter estimation to have been completed on a larger fraction of the candidates. 
We note however that progress is being made in automating detector characterisation and event validation procedures within the Collaboration, for example by \citet{Davis_2026}, which will further reduce the cost of working with these candidates.

We have endeavoured to perform these analyses with a workflow which replicates that used in \ac{GWTC4} as closely as possible, while using tools and data which are available publicly, including the strain data as published by \ac{GWOSC}.
We have, however, made a number of minor compromises to make this achievable, and these are described in section~\ref{sec:differences}.
These changes do not reduce the validity of our results, but for example the lack of results from a second waveform prevents us from studying waveform systematics effects which might be present in the inference.
We have provided a datarelease~\citep{datarelease} which will allow our analyses to be reproduced, and contains the posterior samples for these candidates in the same format as results published from \ac{GWTC4}~\citep{collaborationGWTC40ParameterEstimation2025}.

We identify a small number of candidates, which if of astrophysical origin, stand out as remarkable.
GW230902\_172430 shows evidence of the sources having aligned spins, and in contrast GW230823\_142524 shows evidence of the sources' spins being anti-aligned.
GW230920\_064709 is consistent with being an \ac{NSBH} merger; only a small number of these have been observed to date in comparison to \acp{BBH}.
At least two candidates, GW230902\_122814 and GW231102\_232433, also show evidence of \emph{prior dominance}, which may affect the validity of the inferred parameters for these two candidates, further work may be valuable on better understanding this behaviour for low \ac{SNR} signals.

Despite the candidates we consider in this work being excluded from the set which received full \ac{PE} in \ac{GWTC4}, we demonstrate that real information is contained in the data for the majority of these signals.
The choice of a \ac{far} threshold in \ac{GWTC4} is motivated by the desire to produce a relatively pure sample of results.
This is a pragmatic choice, but we argue that it may be worth revisiting in the future, especially as increased automation and the availability of automated review checks such as those we demonstrated in section~\ref{sec:review} reduce the labour-intensity of producing \ac{PE} samples.
Disagreements over the values of \pastro{} and \ac{far} between the search pipelines, as illustrated in figure~\ref{fig:pipelines}, confound the issue of choosing a more relaxed threshold than the one which was chosen for \ac{GWTC4}, as we do not find that there is a clear correlation between any given pipeline's estimate of these statistics and the ultimate outcome of our attempt to perform \ac{PE}.

As ongoing observations from \ac{gw} detectors continue to expand the catalogue of candidates, it is important that the candidates lurking around the statistical threshold are not entirely forgotten.
These candidates are informative, though require care for a variety of reasons when handling.
The work required to analyse these is often greater than what is required for louder candidates, but improvements in automation provide a viable means of extracting the information from these.

In the future we hope to both develop techniques and workflows which will allow us to exclude candidates (e.g. using~\citealp{Davis_2026}) which are consistent with noise more efficiently, and also ways of producing results using multiple waveform families (e.g. utilising techniques such as those developed in~\citealp{5hvn-3lmh}) in order to allow investigations into effects such as waveform systematics in these low \ac{SNR} candidates.

\section*{Data availability}

The \ASIMOV{} blueprints, configuration files, and posterior samples for this work are available at~\citep{datarelease}.

\section*{Acknowledgements}

DW is supported by Science and Technology Facilities Council (STFC) grant UKRI2487. 
MJW acknowledges partial support from STFC grant UKRI2490.
The authors acknowledge the use of the LLM tool Claude in preparing this manuscript, however all outputs from the LLM were verified by the authors, and text from the model used in this manuscript has been edited as required by the authors.

We would like to thank Ann-Kristin Malz for a number of helpful suggestions on the manuscript.

This research has made use of data or software obtained from the Gravitational Wave Open Science Center (gwosc.org), a service of the LIGO Scientific Collaboration, the Virgo Collaboration, and KAGRA. This material is based upon work supported by NSF's LIGO Laboratory which is a major facility fully funded by the National Science Foundation, as well as the Science and Technology Facilities Council (STFC) of the United Kingdom, the Max-Planck-Society (MPS), and the State of Niedersachsen/Germany for support of the construction of Advanced LIGO and construction and operation of the GEO600 detector. Additional support for Advanced LIGO was provided by the Australian Research Council. Virgo is funded, through the European Gravitational Observatory (EGO), by the French Centre National de Recherche Scientifique (CNRS), the Italian Istituto Nazionale di Fisica Nucleare (INFN) and the Dutch Nikhef, with contributions by institutions from Belgium, Germany, Greece, Hungary, Ireland, Japan, Monaco, Poland, Portugal, Spain. KAGRA is supported by Ministry of Education, Culture, Sports, Science and Technology (MEXT), Japan Society for the Promotion of Science (JSPS) in Japan; National Research Foundation (NRF) and Ministry of Science and ICT (MSIT) in Korea; Academia Sinica (AS) and National Science and Technology Council (NSTC) in Taiwan.

We also acknowledge the use of computing resources provided by the Institute for Gravitational Research which enabled the \ac{PE} analyses presented in this work.

This document has been assigned LIGO document control center number LIGO-P2600589.
%\section*{References}

% arXiv compiles the .bbl without a class that defines this (mnras.bst emits it)
\providecommand{\newblock}{}
\bibliographystyle{mnras}
\bibliography{references}

\appendix

\section{Automated posterior review tools}\label{sec:posterior-tools}

The checks used to review each analysis are implemented in a Python package, \texttt{gw-catalog-tools}~\citep{software-gw-catalog-tools}, which can be run on a single posterior samples file or automatically by\ASIMOV{}, where they are applied to every analysis awaiting review and produce a report for the analyst.
Each check can provide a \emph{failure}, which means the analysis should not be accepted as it is; a \emph{warning}, which flags something for an analyst to examine; or is \emph{informational}, highlighting for example unusual values in the results.
The checks screen the posteriors and do not replace inspecting them: an analysis which they flag may still be approved, and a glitch which gives a plausible posterior could pass them all.
The thresholds were chosen for a catalogue of near-threshold candidates, where a signal has a network \ac{SNR} of roughly 6--15, and were set using the analyses of this work.
The tests are:
\begin{itemize}
    \item \textbf{Railing of chirp mass or luminosity distance} (failure). 
    The posterior of each is histogrammed into 50 bins, and the analysis fails if the bin at either end holds more than 2\% of the density of the fullest bin.
    A new bound is automatically suggested, in proportion to the support at the end, which is how the bounds were widened for the reruns described in section~\ref{sec:review}.
    Railing at the lower end of the distance is ignored if the prior already extends below the suggested bound.
    
    \item \textbf{Multiple modes} (warning). 
    The marginal posteriors of the mass ratio, chirp mass, and the logarithm of the distance are smoothed with a Gaussian kernel, and a mode is a peak at least 8\% as high as the highest and holding at least 3\% of the probability.
    This detects a second mode which holds 20\% or more of the probability, and finds no second mode in samples from Gaussian, log-normal, gamma, or bounded distributions.
    A shoulder against a prior bound is not a mode; the railing check reports it.
    
    \item \textbf{Spins against the prior} (warning). 
    For $a_1$, $a_2$, and $\chi_p$, whether the fraction of samples above 0.7 exceeds that of the prior by more than 0.1, and for $\chi_\mathrm{eff}$ the same beyond $\pm 0.9$.
    A posterior pushed against the spin bounds is characteristic of a glitch or of an analysis dominated by noise.
    
    \item \textbf{Signal plausibility} (warning). Three conditions.
    (i) The analysis is consistent with noise if the natural logarithm of the Bayes factor is below 1 or the median network \ac{SNR} is below 4.
    (ii) The signal is implausibly loud if the median network \ac{SNR} exceeds 20, which for a candidate below the significance threshold suggests that the analysis has fitted a loud noise transient.
    (iii) A single detector dominates if it holds more than 75\% of the squared \ac{SNR}.
    \item \textbf{Consistency of the derived parameters} (failure). The total mass, mass ratio, and chirp mass agree with the component masses to within 0.1\%.
    \item \textbf{Likelihood} (failure). The log-likelihood and log-prior of every sample are finite, and the log-likelihood is not constant.
    \item \textbf{Network \ac{SNR}} (failure). The network optimal \ac{SNR} should equal the sum in quadrature of those of the detectors, to within 1\%.
    \item \textbf{Sky area} (informational). The 90\% credible area on a grid of equal-area cells, which is thousands of square degrees for all of the candidates in this work.
\end{itemize}
Two further tests of the analysis configurations and diagnostics provided by the sampler, rather than of the posteriors, verify that no samples were lost from the sampler output, and that the waveform arguments in the configuration of each analysis are those intended.
Table~\ref{tab:matrix} shows which of these tests flag the candidates which could not be approved.

% \todo{Describe the remaining tools: a per-analysis summary of posterior percentiles and prior bounds, the comparison of the prior used between consecutive analyses, and the Q-transform scans of the strain data around each candidate (\texttt{scripts/qscans.py}).}

% \section{Analysis history and posterior summaries}\label{sec:history}

% Figure~\ref{fig:history} shows the sequence of analyses of each event and the outcome of the review of each, and Table~\ref{tab:summary} gives the medians and 90\% credible intervals of the parameters of the events with an approved analysis.

% \begin{figure*}
%     \centering
%     \includegraphics[width=\linewidth]{figures/history.pdf}
%     \caption{The IMRPhenomXPHM-SpinTaylor analyses of each event, in the order in which they were prepared, and the outcome of the review of each. The symbol on the left of each row is the outcome of the event, as in Figure~\ref{fig:selection}. Reruns which repeated an earlier prior because of an error in the blueprint (Appendix~\ref{sec:asimov}) are omitted, as are analyses which were cancelled before they completed, and the analyses are numbered in the order of the remaining ones. An analysis is not finished if it has not completed, and awaiting review if it has completed but has not been reviewed.}
%     \label{fig:history}
% \end{figure*}

\end{document}